\documentclass[showpacs,prl,floatfix,preprintnumbers,twocolumn]{revtex4}
\usepackage{graphicx,url,amssymb,amsmath,longtable,rotating,color,units,wasysym,subfigure,listings,bm, epstopdf, bigints}

\begin{document}
\title{A Systematic Study of the Stochastic Gravitational-Wave Background due to Stellar Core Collapse}
\author{K. Crocker$^{a}$, T. Prestegard$^a$, V. Mandic$^a$\footnote{mandic@physics.umn.edu}, T. Regimbau$^b$, K. Olive$^{c,a}$,  E. Vangioni$^d$}
\affiliation{$^a$ School of Physics and Astronomy, University of Minnesota, Minneapolis, MN 55455, USA\\
$^b$ Departement Artemis, Observatoire de la C\^ote d'Azur, CNRS, F-06304 Nice,  France\\
$^c$ William I. Fine Theoretical Physics Institute, University of Minnesota, Minneapolis, MN 55455, USA\\
$^d$ Sorbonne Universit\'es, UPMC Univ. Paris 6 et CNRS, UMR 7095, Institut d'Astrophysique de Paris, 98 bis bd Arago, 75014 Paris, France}
\date{\today}

\begin{abstract}
Stellar core collapse events are expected to produce gravitational waves via several mechanisms, most of which are not yet fully understood due to the current limitations in the numerical simulations of these events.
In this paper, we begin with an empirical functional form that fits the gravitational-wave spectra from existing simulations of stellar core collapse and integrate over all collapse events in the universe to estimate the resulting stochastic gravitational-wave background.
We then use a Gaussian functional form to separately fit and model a low-frequency peak in the core-collapse strain spectra, which likely occurs due to prompt convection.
We systematically study the parameter space of both models, as well as the combined case, and investigate their detectability by upcoming gravitational-wave detectors, such as Advanced LIGO and Einstein Telescope.
Assuming realistic formation rates for progenitors of core-collapse supernovae, our results indicate that both models are 2--4 orders of magnitude below the expected sensitivity of Advanced LIGO, and 1--2 orders of magnitude below that of the Einstein Telescope.
\end{abstract}

\maketitle

\section{Introduction}\label{sec:intro}
The superposition of gravitational waves (GWs) generated by many uncorrelated and unresolved sources in the universe leads to a stochastic gravitational-wave background (SGWB).
Such a background could be of cosmological or astrophysical origin.
A cosmological SGWB may be produced by cosmic (super)strings \cite{caldwellallen,DV1,DV2,cosmstrpaper,olmez1,olmez2}, inflation~\cite{grishchuk,starob,eastherlim,peloso}, primordial black hole binaries \cite{bird, sasaki, mandic_pbbh}, alternative cosmologies~\cite{PBB1,PBBpaper}, and a variety of other phenomena.
Astrophysical SGWB models
%see \cite{regimbau} for a review) integrate GW signals from astrophysical objects across the universe,
include coalescences of compact objects in binary systems, like binary neutron stars or binary black holes \cite{phinney,kosenko,2006ApJ...642..455R,zhu,2011PhRvD..84h4004R,2011PhRvD..84l4037M,StochCBC,2013MNRAS.431..882Z,2013MPLA...2850174E,2015A&A...574A..58K,2015MNRAS.449.2700E}, rotating neutron stars \cite{RegPac,owen,1999MNRAS.303..258F,barmodes1,barmodes2,barmodes3,2004MNRAS.351.1237H,2011ApJ...729...59Z,2012PhRvD..86j4007R,2013PhRvD..87f3004L}, magnetars \cite{cutler,2006A&A...447....1R,RegMan,2011MNRAS.411.2549M,2011MNRAS.410.2123H,2013PhRvD..87d2002W,2013PhRvD..87f3004L}, the first stars \cite{firststars}, and white dwarf binaries~\cite{phinney_whitedwarfs}.

Some of the proposed models have been constrained \cite{cosmstrpaper,StochCBC,paramest,parviol} by searches for the isotropic \cite{S3stoch,S4stoch,S5stoch,S6stoch,H1H2stoch} and anisotropic SGWB \cite{S4radiometer,S5SPH}.
These searches have established upper limits on the energy density in the SGWB using data acquired by the first generation interferometric gravitational-wave detectors LIGO \cite{LIGOS1,LIGOS5} and Virgo \cite{Virgo1}.
The second-generation Advanced LIGO (aLIGO) detectors have recently conducted their first observational run \cite{GW150914det,aLIGO2}, detecting the first gravitational-wave signals due to mergers of binary black hole systems \cite{GW150914,GW151226}, and producing new estimates of the SGWB due to binary black hole mergers \cite{GW150914stoch}.
Other second-generation detectors, including Advanced Virgo \cite{aVirgo}, GEO-HF \cite{GEOHF,GEOHF2}, and KAGRA \cite{KAGRA1,KAGRA2}, are expected to become operational in the coming years.
The third-generation gravitational-wave detectors, including the Einstein Telescope~\cite{ET} and Cosmic Explorer~\cite{CosmExpl}, are also currently being conceptualized.
%The Einstein Telescope, for which the design study was completed in Europe \cite{ET}, is an example of such a detector.

%A promising source of gravitational waves (in terms of accessibility to second and third-generation detectors) is the process of stellar core collapse.
Gravitational waves from stellar core collapse are expected to be produced via several mechanisms, including a quasi-periodic signal generated during the post-shock convection phase, hot-bubble convection and the standing accretion shock instability (SASI)~\cite{murphy2009, marek2009, kotake2009}, anisotropic neutrino emission~\cite{marek2009, kotake2009, ott2009}, and the ringdown of the potentially newly formed black hole.
Even though full three-dimensional simulations that include a complete set of relevant physical processes are not yet computationally feasible, predictions have been made about the gravitational-wave signals emitted during core collapse \cite{OttPRL,ott2013,muller2013,yakunin}. These predictions have been used to compute estimates of the corresponding SGWB from stellar core collapse due to both standard and early (Population III) stars \cite{ferrari_ccbh,2000PhRvD..61l4015D,2002MNRAS.330..651D,araujo,2004MNRAS.348.1373D,2004CQGra..21S.545D,buonanno,coward,2010MNRAS.409L.132Z,marassi_cc,firststars,pacucci}. However, the dependency of the gravitational-wave signal on stellar progenitor properties, such as mass or spin, is not well known, and the rate of core collapse events is similarly uncertain. This means that the resulting SGWB estimates are necessarily only approximate.

In this paper, we try to take a more systematic look at the stellar core collapse SGWB model. We begin with an empirical functional form that has been used in the past to describe the gravitational-wave spectrum emitted by a single stellar core collapse. We fit this functional form to the available GW spectra obtained in numerical simulations of the core collapse process to determine the plausible range of the free parameters. In the process of studying these simulations, we observe that many simulations predict a low-frequency peak that our functional form is unable to capture. As a result, we model this feature using a second functional form.
For both of these models, we compute the SGWB due to all stellar core collapse events in the universe, assuming that the redshift distribution of core collapse events follows the star formation rate. We use the star formation rate model from \cite{vangioni}, which is shown to agree with observations of metallicity, optical depth, and the reionization redshift, as obtained from CMB measurements. We then scan the parameter space and determine the regions that will be accessible to Advanced LIGO (at design sensitivity) and Einstein Telescope detectors as the representative second and third generation detectors, respectively. We note that this study is a follow up on the study presented in \cite{crocker}, which focused only on gravitational waves produced by the ringdown of the black hole formed at the end of the core collapse.

In Section 2, we present a model of the SGWB due to stellar core collapse, identify the plausible ranges of the model parameter space, and discuss the low-frequency feature in the spectrum. In Section 3, we perform a systematic scan of the parameter space of this SGWB model and discuss its accessibility to the future detectors. We summarize our results in Section 4.

\section{SGWB Model due to Stellar Core Collapse}\label{sec:astro_SGWB}
Following \cite{regimbau,RegMan,StochCBC,GW150914stoch}, we define the GW energy density normalized by the critical energy density:
\begin{eqnarray}
\Omega_{\rm GW}(f) & = & \frac{1}{\rho_c } \frac{d\rho_{\rm GW}}{d \ln f},
\end{eqnarray}
where $\rho_{\rm GW}(f)$ is the energy density in gravitational waves in the frequency band $(f,f+df)$. The critical energy density is the energy density necessary to close the universe, $\rho_c = \frac{3 H_0^2 c^2}{8\pi G}$, where $H_0$ is the current value of the Hubble constant, taken to be 67.7 km/s/Mpc, $G$ is the gravitational constant, and $c$ is the speed of light. Then, following \cite{GW150914stoch}, we have:
\begin{eqnarray}
\Omega_{\rm GW}(f) & = & \frac{f}{\rho_c H_0} \int_0^{z_{\max}} \frac{R(z)\frac{dE}{df_e}(f_e) dz}{(1+z) E(\Omega_{\text m}, \Omega_{\Lambda},z)},
\label{eq:omega}
\end{eqnarray}
where the energy spectrum emitted by a single astrophysical source is denoted $dE/df_e (f_e)$ and it is evaluated at the source (emitted) frequency $f_e$ which is related to the observed frequency as $f_e = f(1+z)$. The rate of stellar core collapse events is denoted $R(z)$ and it can be expressed in terms of the star formation rate (SFR) $R_*(z)$:
\begin{equation}
R(z) = \lambda_\text{CC} R_*(z),
\label{eqrv}
\end{equation}
where $\lambda_\text{CC}$ denotes the mass fraction of stars which undergo the core-collapse process.
In our analysis, we will treat $\lambda_\text{CC}$ as a free parameter, but we note that it can be estimated from the initial mass function (IMF).
Assuming the Salpeter IMF $\phi(m) = N m^{-2.35}$ normalized such that $\int_{0.1 M_{\odot}}^{\infty} \phi(m) \, m \, dm = 1$, we can estimate $\lambda_\text{CC}$ by assuming that stars with masses larger than $8 M_{\odot}$ undergo core collapse: $\lambda_\text{CC} = \int_{8 M_{\odot}}^{\infty} \phi(m) dm \approx 0.007 M_{\odot}^{-1}$.

The factor of $1+z$ in Eq.~\ref{eq:omega} corrects for the cosmic expansion, and converts the time in the source frame to the detector frame. Finally, $E(\Omega_\text{m},\Omega_{\Lambda},z)$ in Eq.~\ref{eq:omega} captures the redshift dependence of the comoving volume:
\begin{eqnarray}
E(\Omega_\text{m},\Omega_{\Lambda},z) = \sqrt{\Omega_\text{m}(1+z)^3 + \Omega_{\Lambda}},
\end{eqnarray}
where $\Omega_\text{m} = 0.309$ and $\Omega_{\Lambda} = 0.691$ denote, respectively, the energy density in matter and in dark energy~\cite{planck2015_cp}. Combining the above, we have
\begin{eqnarray}
\Omega_{\rm GW}(f) & = & \frac{8\pi Gf\lambda_\text{CC}}{3H_0^3 c^2} \int dz \frac{ R_*(z)}{(1+z)E(\Omega_\text{m},\Omega_{\Lambda},z)} \; \frac{dE}{df_e}(f_e).\nonumber \\
&&
\label{eqomega}
\end{eqnarray}

Finally, we comment on our choice of the star formation rate model.
The SFR and its redshift dependence have been examined by multiple authors, who have proposed several functional forms~\cite{lilly,hopkins,fardal,wilkins,nagamine,springel,vangioni}.
%We use the functional form proposed by \cite{vangioni} that is based on the gamma ray burst (GRB) observations at high redshifts ~\cite{robertsonellis,wang,kistler}---specifically, we use the GRB rate of \cite{kistler} based on the normalization from \cite{trenti,behroozisilk}.
We use the functional form proposed by \cite{vangioni} that is based on measurements of the luminosity function in high-redshift galaxies~\cite{behroozi}.
As noted in \cite{vangioni}, this SFR model is consistent with metallicity observations, as well as with the optical depth and reionization redshift inferred from cosmic microwave background observations~\cite{WMAP,planck2015_cp}.
This SFR model is given by the Springel \& Hernquist functional form~\cite{springel}:
\begin{eqnarray}
R_*(z) = \nu \; \frac{p e^{q(z-z_m)}}{p-q+q e^{p(z-z_m)}}
\end{eqnarray}
%
%with the following parameters \cite{vangioni}: $\nu = 0.146$ M$_{\odot}/{\rm yr}/{\rm Mpc}^3 $, $z_m = 1.72$, $p = 2.80$, and $q = 2.46$.
%We note that at low redshifts, this SFR model agrees well with an SFR model based on luminosity measurements \cite{vangioni}.%, as shown in Figure \ref{sfr}.
with the following parameters \cite{vangioni}: $\nu = 0.178$ M$_{\odot}/{\rm yr}/{\rm Mpc}^3 $, $z_m = 2.00$, $p = 2.37$, and $q = 1.80$.
We note that at low redshifts, this SFR model agrees well with a model based on GRB observations at high redshift~\cite{vangioni}.%, as shown in Figure \ref{sfr}.
At high redshifts, the difference is more substantial, but as noted in \cite{crocker}, the contribution to the SGWB from high-redshift stellar core collapse events is not dominant, and therefore the choice between these two SFR models does not make a significant difference in the resulting SGWB.
As a result, we continue our analysis assuming the luminosity function-based SFR model, since it matches CMB observations more closely in terms of optical depth and redshift of reionization.
%As a result, we continue our analysis assuming the GRB-based SFR model.

\subsection{High Frequency Model}
As noted above, the physics of the stellar core collapse process is not yet fully understood, and three-dimensional numerical simulations which include the complete physical description of the process are yet to be conducted. Past simulations have primarily been in two dimensions \cite{muller2013}; although some are three-dimensional, these simulations typically use a coarser physical description \cite{ott2013}. In \cite{buonanno,firststars}, it was shown that the following functional form could describe the GW amplitude spectra $\tilde{h}(f_e)$ emitted during the core collapse process (in the local frame of the star):
\begin{eqnarray} \label{high_freq_strain}
f_e |\tilde{h}(f_e)| = \frac{G}{\pi c^4 D} E_{\nu} \langle q \rangle \left( 1 + \frac{f_e}{a} \right)^3 e^{-f_e/b},
\label{funcform}
\end{eqnarray}
where $a$ and $b$ are free parameters of the model, $D$ is the distance to the star (assumed small enough that redshifting effects can be ignored), $E_{\nu}$ is the energy carried away by neutrinos during the core collapse and $\langle q \rangle$ is the luminosity-weighted averaged neutrino anisotropy \cite{buonanno}.
%The source distance is denoted $D$, and is small enough that the redshift is negligible.
%The frequencies are therefore denoted $f_e$, the emitted frequency in the source frame.

The GW energy spectrum from a single core collapse event can then be computed as:
\begin{eqnarray}\label{eq:dEdf_highF}
\frac{dE}{df_e}(f_e) & = & \frac{\pi^2 c^3 D^2}{G} f_e^2 |h(f_e)|^2 \nonumber \\
& = & \frac{G}{c^5} E_{\nu}^2 \langle q \rangle^2 \left( 1 + \frac{f_e}{a} \right)^6 e^{-2f_e/b}.
\end{eqnarray}

\begin{figure}
  \includegraphics[width=3.2in]{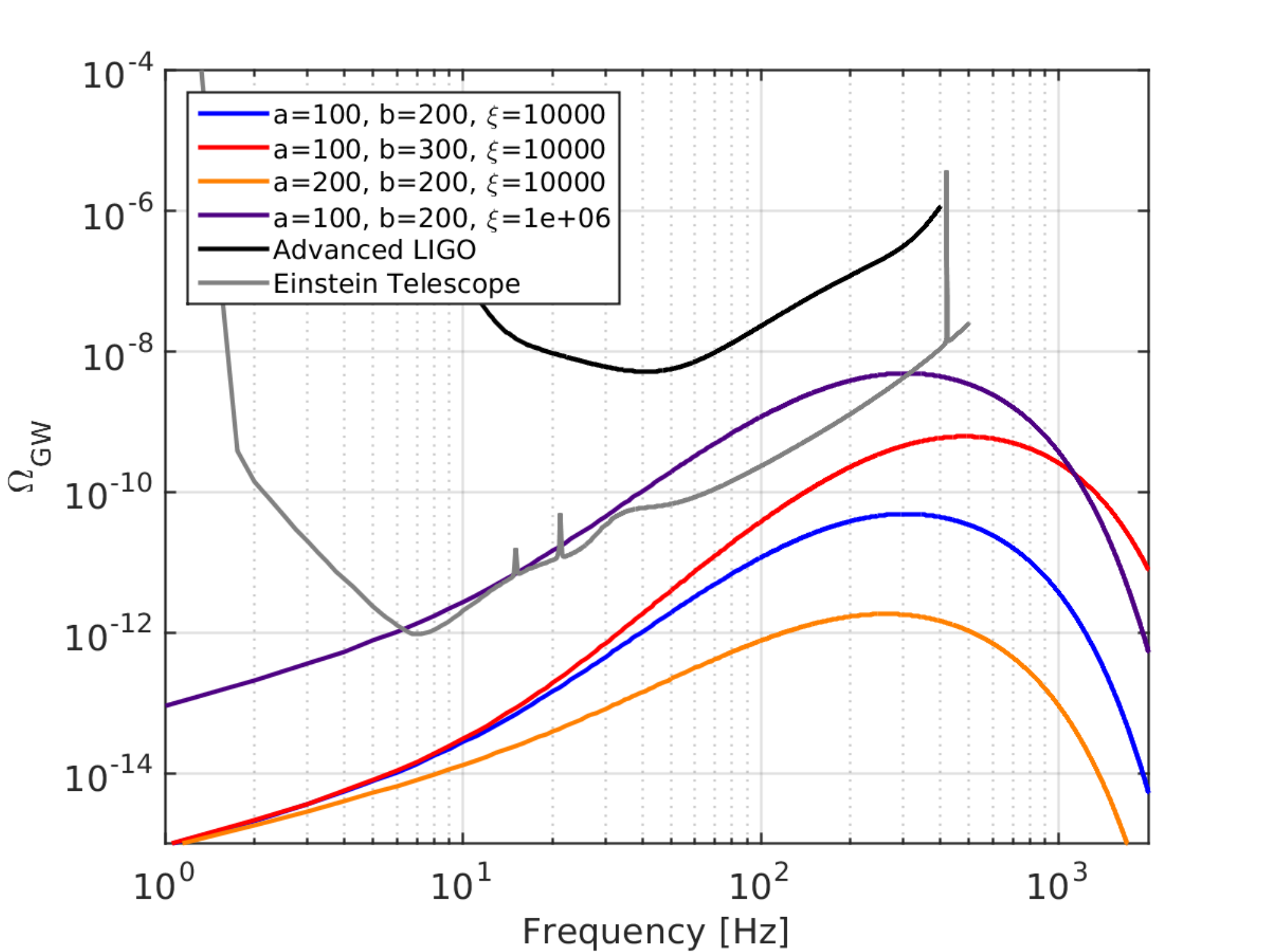}
  \caption{$\Omega_\text{GW}(f)$ for various parameter choices for the high-frequency model of an SGWB produced by stellar core-collapse. $a$ and $b$ are in units of Hz, while $\xi$ has units of m$^2$/s. Overall, this model has three free parameters: $a$, $b$, and $\xi$. In general, the GW energy density increases as $a$ decreases; at frequencies above a few Hz, it essentially scales proportionally to $a^{-6}$. As $b$ increases, the rate at which the exponential term in Eq.~\ref{eq:high} suppresses the GW energy density is decreased; this leads to an increase in the overall energy density and pushes the peak of the distribution to higher frequencies. Finally, the GW energy density scales proportionally to $\xi$. The effect of each parameter on the spectrum is further illustrated in Figure~\ref{Omega_vs_ab}. Also shown are the $\text{SNR} = 2$ sensitivities of Advanced LIGO \cite{aLIGOsens} and ET \cite{ET}, assuming 1 year of exposure and two colocated detectors.}
  \label{high_frequency}
\end{figure}

Inserting this spectrum into Eq.~\ref{eqomega}, we obtain the explicit form of the SGWB spectrum:
%\begin{eqnarray} \label{eq:high}
%{\setlength{\abovedisplayskip}{0pt}
%& \Omega_{\rm GW}(f) = \frac{8\pi Gf\xi}{3H_0^3 c^2} \nonumber \\
%& \bigintssss dz \frac{ R_*(z)}{(1+z) E(\Omega_\text{m},\Omega_{\Lambda},z)} \; \left( 1 + \frac{f(1+z)}{a} \right)^6 e^{-2f(1+z)/b}, \\
%& \xi = \frac{G \lambda_{\rm CC}}{c^5} E_{\nu}^2 \langle q \rangle^2
%}
%\end{eqnarray}
\begin{align}\label{eq:high}
  \begin{split}
    \Omega_{\rm GW}(f) &= \frac{8\pi Gf\xi}{3H_0^3 c^2} \bigintssss dz \, \frac{ R_*(z)}{(1+z) E(\Omega_\text{m},\Omega_{\Lambda},z)} \\
    &\phantom{=} \left( 1 + \frac{f(1+z)}{a} \right)^6 e^{-2f(1+z)/b},
  \end{split}
\end{align}
where $\xi$ is a combination of unknown scaling factors, defined as
\begin{equation}
\xi = \frac{G \lambda_{\rm CC}}{c^5} E_{\nu}^2 \langle q \rangle^2
\end{equation}
In anticipation of the low-frequency peak which is discussed and modeled below, we will refer to this as the high-frequency model. Figure \ref{high_frequency} shows examples of spectra generated with this model. See the caption for discussion of the effect of the choice of model parameters on the morphology  of the spectrum.

Figure \ref{high_frequency} also shows the SGWB sensitivities of the Advanced LIGO \cite{aLIGOsens} and Einstein Telescope \cite{ET} detectors, assuming the detection statistic defined by \cite{allen-romano} for which the signal to noise ratio is defined as:
\begin{equation}
  \text{SNR} =\frac{3 H_0^2}{10 \pi^2} \sqrt{2T} \left[
\int_0^\infty df\>
\frac{\gamma^2(f)\Omega_{\rm GW}^2(f)}{f^6 P_1(f) P_2(f)} \right]^{1/2}\,,
\label{snr}
\end{equation}
where $T$ is the observation time (set to 1 year in our case), $\gamma(f)$ is the overlap reduction function for the chosen detector pair (set to 1, assuming colocated detector pairs) arising from the different locations and orientations of the detectors \cite{allen-romano}, and $P_1(f)$ and $P_2(f)$ are the strain power spectral densities of the two detectors. In Figure \ref{high_frequency}, we plot the $\text{SNR} = 2$ curves.

\begin{figure*}
  \centering
  \includegraphics[width=3.2in]{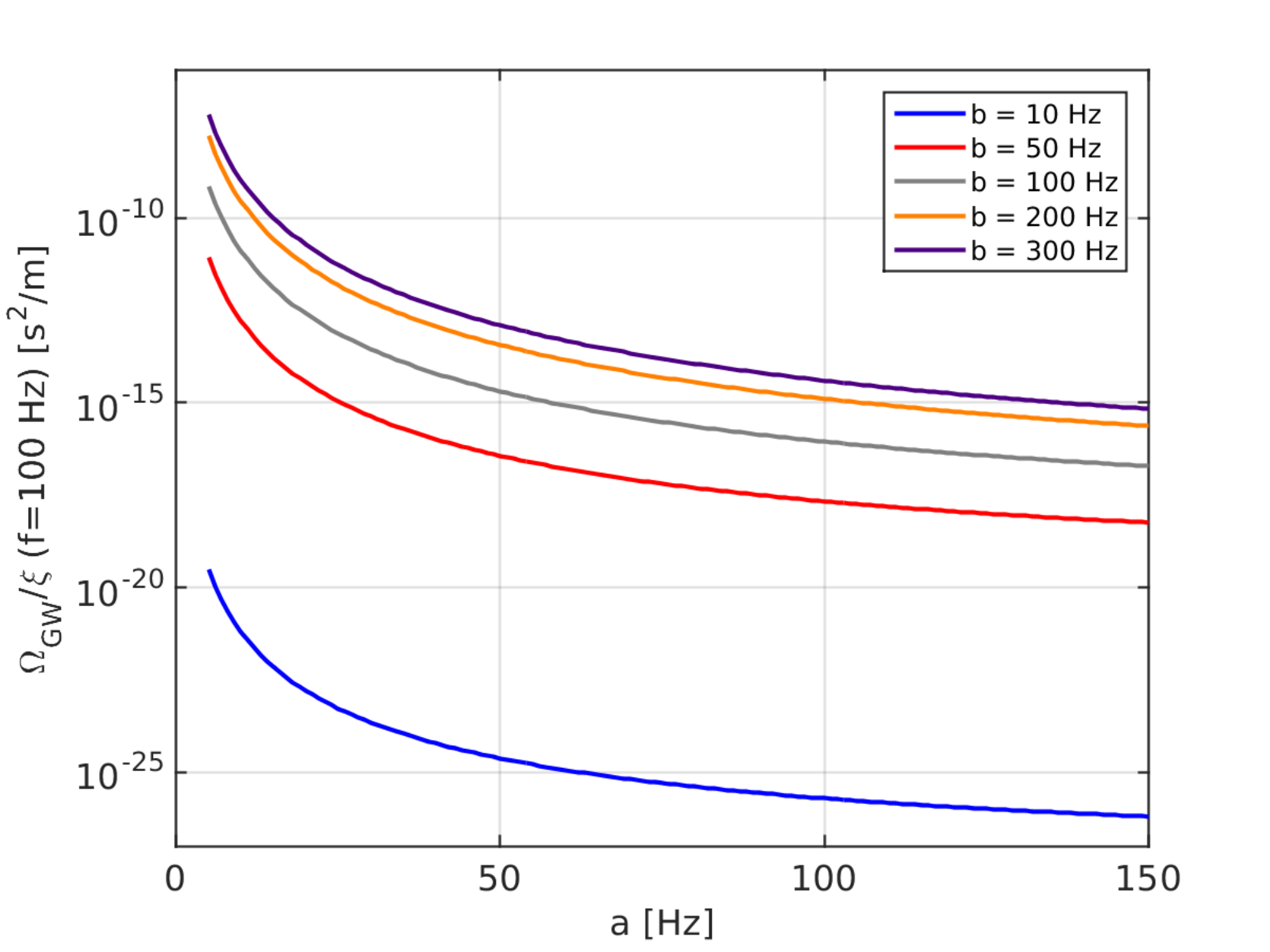} \quad
  \includegraphics[width=3.2in]{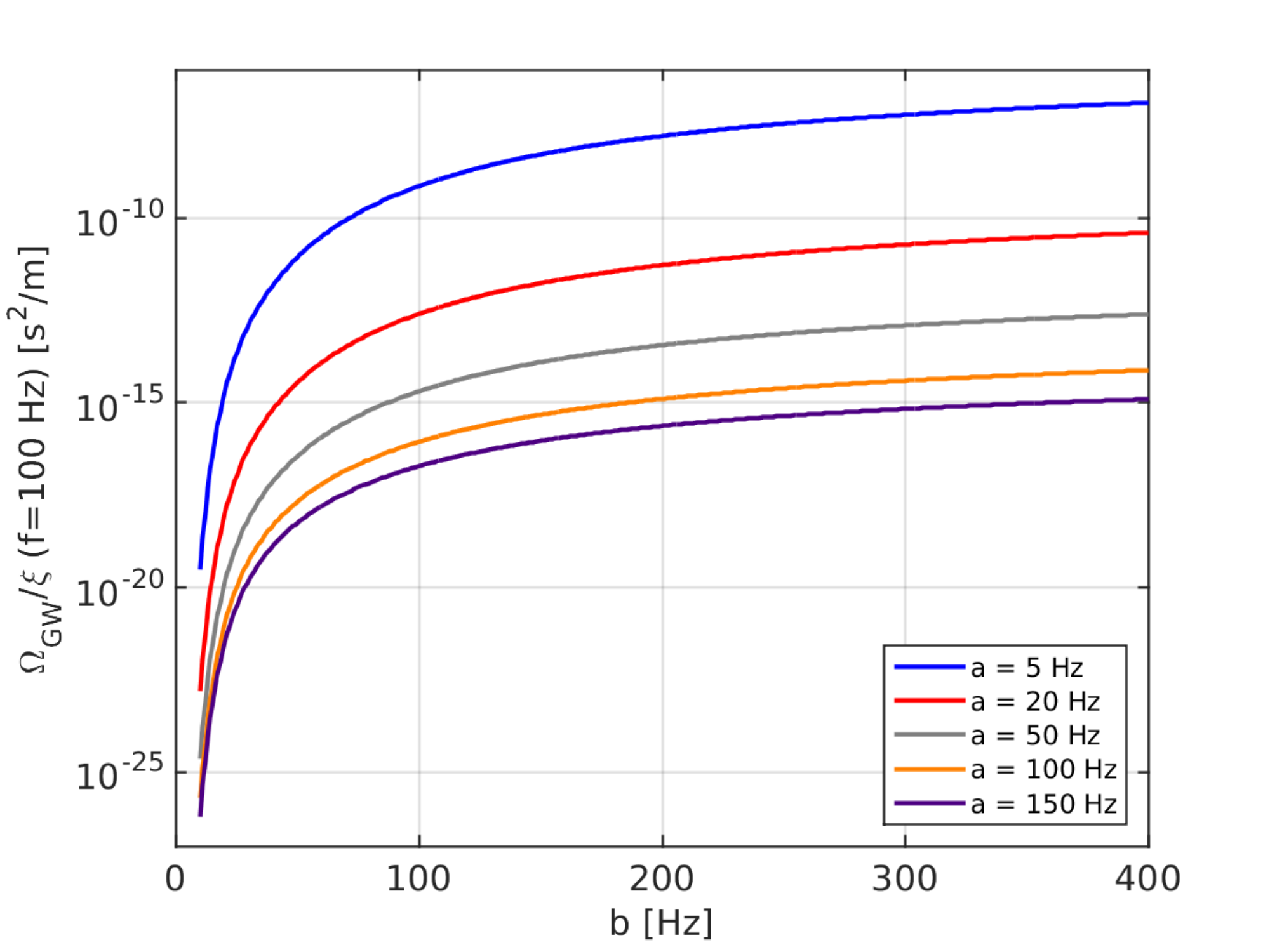}
  \caption{Left: $\Omega_\text{GW}/\xi$ at 100 Hz versus $a$ for several choices of $b$. Right: $\Omega_\text{GW}/\xi$ at 100 Hz versus $b$ for several choices of $a$. Overall, $\Omega_\text{GW}$ tends to increase as $a$ decreases and as $b$ increases. See the text for more details.}
  \label{Omega_vs_ab}
\end{figure*}

In Figure~\ref{Omega_vs_ab}, we show the effect of $a$ and $b$ on the spectrum by plotting $\Omega_\text{GW}/\xi$ at 100 Hz as a function of these parameters.
For a fixed value of $b$, increasing $a$ leads to decreased $\Omega_\text{GW}$; this is apparent from the fact that $\Omega_\text{GW}$ goes approximately as $a^{-6}$ (see Eq.~\ref{eq:high}).
The converse is true for $a$: fixing $a$ and allowing $b$ to grow leads to higher $\Omega_\text{GW}$, since increasing $b$ pushes the exponential term in Eq.~\ref{eq:high} closer to 1.

Similar results were found in \cite{firststars} and \cite{dvorkin}.
In \cite{firststars}, two modes of star formation were considered:
a normal mode of star formation as considered here, and several
possibilities for an additional population of massive stars (Population III)
which were deemed necessary as initial reports of the optical depth were quite high.
Using $a = 200$ and $b=300$, they found a peak value of $\Omega_{\rm GW} h^2
 = 3-4 \times 10^{-10}$ at $f \approx 300$ Hz. In general, the massive
modes made a relatively small contribution (no more than a factor of 30)
at lower frequencies ($f < 100$ Hz). Above 300 Hz, all models were indistinguishable.
More recently, using the same SFR considered here, Ref. \cite{dvorkin}
compared different models of black hole formation from single star collapse and mergers.
Single star collapse also showed peak values of $\Omega_{\rm GW} h^2$ between
$10^{-10}$ and $10^{-9}$, though the peak frequency was model dependent.

In order to perform a scan of the parameter space of this model, it is necessary to determine the plausible ranges of the parameters $a$ and $b$. Since these are empirical parameters, we determine the ranges of their values by fitting the functional form from Eq.~\ref{funcform} to the spectra generated in various simulations of the core collapse. We performed fits systematically to all waveforms provided by \cite{ott2013,abd2014,yakunin,muller2013} and show below several examples to illustrate the quality of the fits and the range of the fitted parameters. We limit the fits to the frequencies below $f_e \approx 400$ Hz: since the SGWB will be dominated by progenitors at redshifts $z \lesssim 3$, imposing the 400 Hz cut-off in the emitted frequency will impact the spectrum only at observed frequencies $\gtrsim 100$ Hz, where the sensitivity of terrestrial GW detectors is significantly reduced (c.f. Figure \ref{high_frequency}).

\begin{figure}
  \includegraphics[width=3.2in]{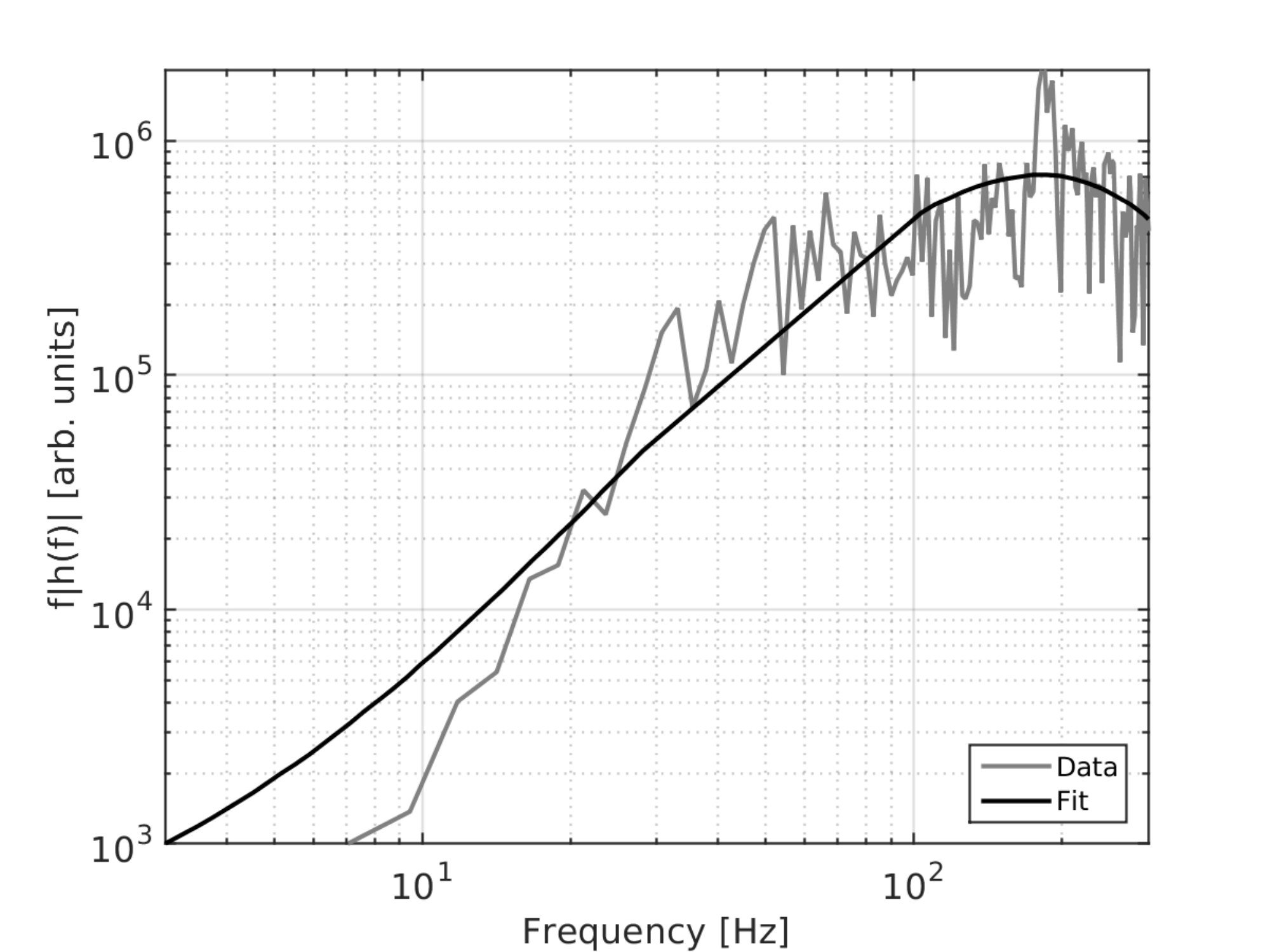}
  \caption{Gravitational-wave signal from the Ott et al. \texttt{s27f$_\text{heat}$1.05} simulation~\cite{ott2013}. In particular, this signal is from a polar observation of a cross-polarized gravitational wave. The original $f|\tilde{h}(f)|$ data is plotted in gray and the fit to this data (with $a = 5$ Hz and $b = 63$ Hz) of our $f|\tilde{h}(f)|$ model (Eq.~\ref{funcform}) is shown in black.
      }
  \label{no_lowpeak_ott}
\end{figure}

The first set of simulated spectra comes from a study performed by Ott et al. \cite{ott2013}. They perform multiple simulations in which a $27 {\rm \; M_{\odot}}$ star is made to undergo core collapse and examine the post-core bounce phase. These simulations are carried out in three dimensions, use general relativity, and apply a neutrino leakage scheme that is consistent with radiation-hydrodynamics simulations. The neutrino heating rate and simulation duration are varied in each simulation  \cite{ott2013} (see Table 1 in \cite{ott2013} for details regarding each simulation). Figure \ref{no_lowpeak_ott} shows $f|\tilde{h}(f)|$ from a cross-polarized gravitational-wave signal, along with a fit of our functional form. This signal was generated from a 184 ms simulation which used a neutrino heating parameter of 1.00 (see \cite{ott2013} for more detail).

\begin{figure}
  \includegraphics[width=3.2in]{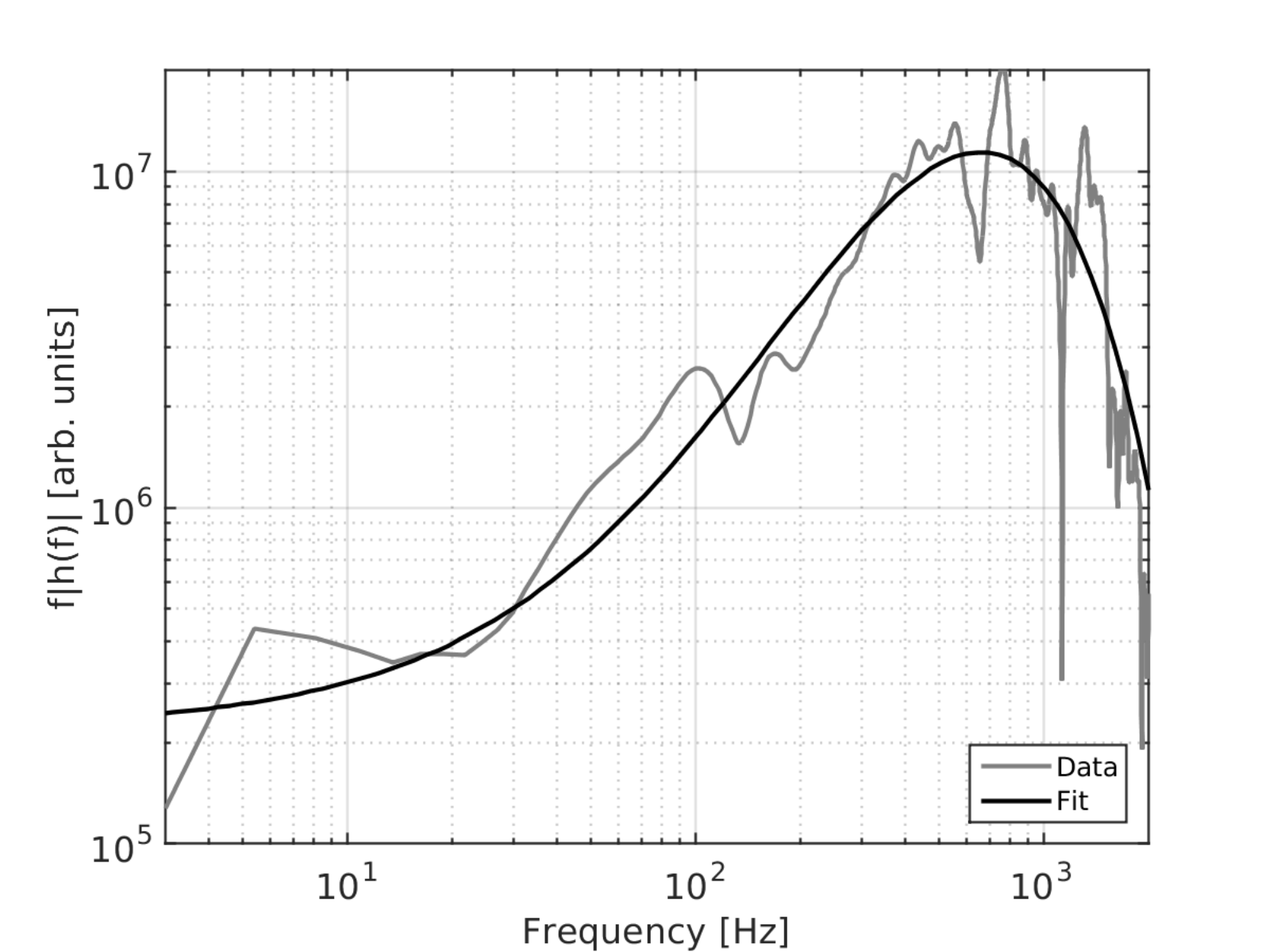}
  \caption{Plus-polarized gravitational wave signal from an equatorial observation computed in the Abdikamalov et al. \texttt{A4O06.5} simulation~\cite{abd2014}. The original $f|\tilde{h}(f)|$ data is plotted in gray, and the fit to this data (with $a = 82$ Hz and $b = 248$ Hz) of our $f|\tilde{h}(f)|$ model (Eq.~\ref{funcform}) is shown in black.
      }
  \label{no_lowpeak_abd}
\end{figure}

We also use simulations performed by Abdikamalov et al. \cite{abd2014}, which examine the angular momentum dependence of GW signals coming from stellar core collapse.
In particular, the collapse of a $12 {\rm \; M_{\odot}}$ progenitor model described in \cite{woosley2007} is carried out in two dimensions under the assumption of axisymmetry and with the use of the general relativistic hydrodynamics code \texttt{CoCoNuT}~\cite{muller2010}.
The simulation also uses an approximate electron capture model, the Lattimer-Swesty equation of state (EOS)~\cite{lattimer-swesty} with bulk modulus $K = 220$ MeV, and a neutrino leakage scheme for postbounce evolution.
The simulation is carried out many times, with the angular momentum in the core varied in each iteration (see Tables I and II in \cite{abd2014} for details).
Figure~\ref{no_lowpeak_abd} shows the strain spectrum from a plus-polarized gravitational wave signal, rescaled by observer distance $D$, for simulation \texttt{A4O06.5} (see Table II in \cite{abd2014}), along with our fit.

\begin{figure}
  \includegraphics[width=3.2in]{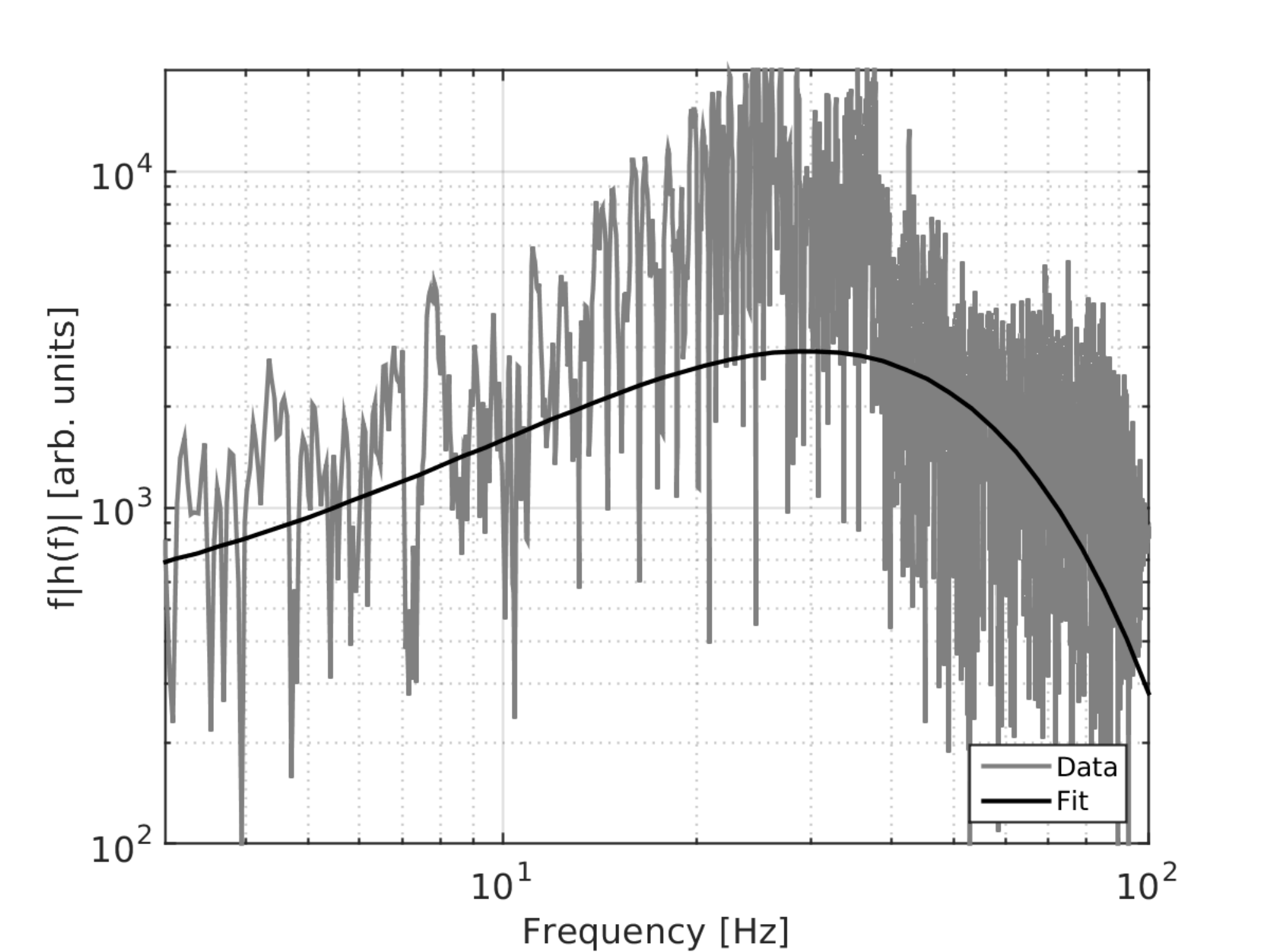}
  \caption{Plus-polarized gravitational wave signal from the Yakunin et al. \texttt{B20-WH07} simulation~\cite{yakunin} produced by changing mass quadrupole moment (as opposed to neutrino emission) is plotted. The original $f|\tilde{h}(f)|$ data is plotted in gray, and the fit of our $f|\tilde{h}(f)|$ model (with $a = 6$ Hz and $b = 11$ Hz, Eq.~\ref{funcform}) is plotted in black.
      }
  \label{no_lowpeak_yak}
\end{figure}

Next, we use spectra from simulations performed by Yakunin et al.~\cite{yakunin}.
These simulations use the \texttt{CHIMERA} code~\cite{chimera} to study the collapse of four Woosley-Heger non-rotating stellar progenitors between 12 M$_{\odot}$ and 25 M$_{\odot}$~\cite{woosley2007}.
These simulations are in two dimensions, use multifrequency neutrino transport in ray-by-ray approximation with relativistic corrections, and a Lattimer-Swesty EOS~\cite{lattimer-swesty} with $K = 220$ MeV \cite{yakunin}.
Figure~\ref{no_lowpeak_yak} shows the strain spectrum from this simulation carried out for the 20 M$_{\odot}$ progenitor, as well as our fit to these data.

%[FROM YAKUNIN ABSTRACT: "All models employ multifrequency neutrino transport in the ray-by-ray approximation, state-of-the-art weak interaction physics, relativistic transport corrections such as the gravitational redshift of neutrinos, two-dimensional hydrodynamics with the commensurate relativistic corrections, Newtonian self-gravity with a general relativistic monopole correction, and the Lattimer–Swesty equation of state with 220 MeV compressibility, and begin with the most recent Woosley–Heger nonrotating progenitors in this mass range." IS ANY MORE OF THIS RELEVANT? \commentt{Tanner}{Not in my opinion.}]

\begin{figure}
  \includegraphics[width=3.2in]{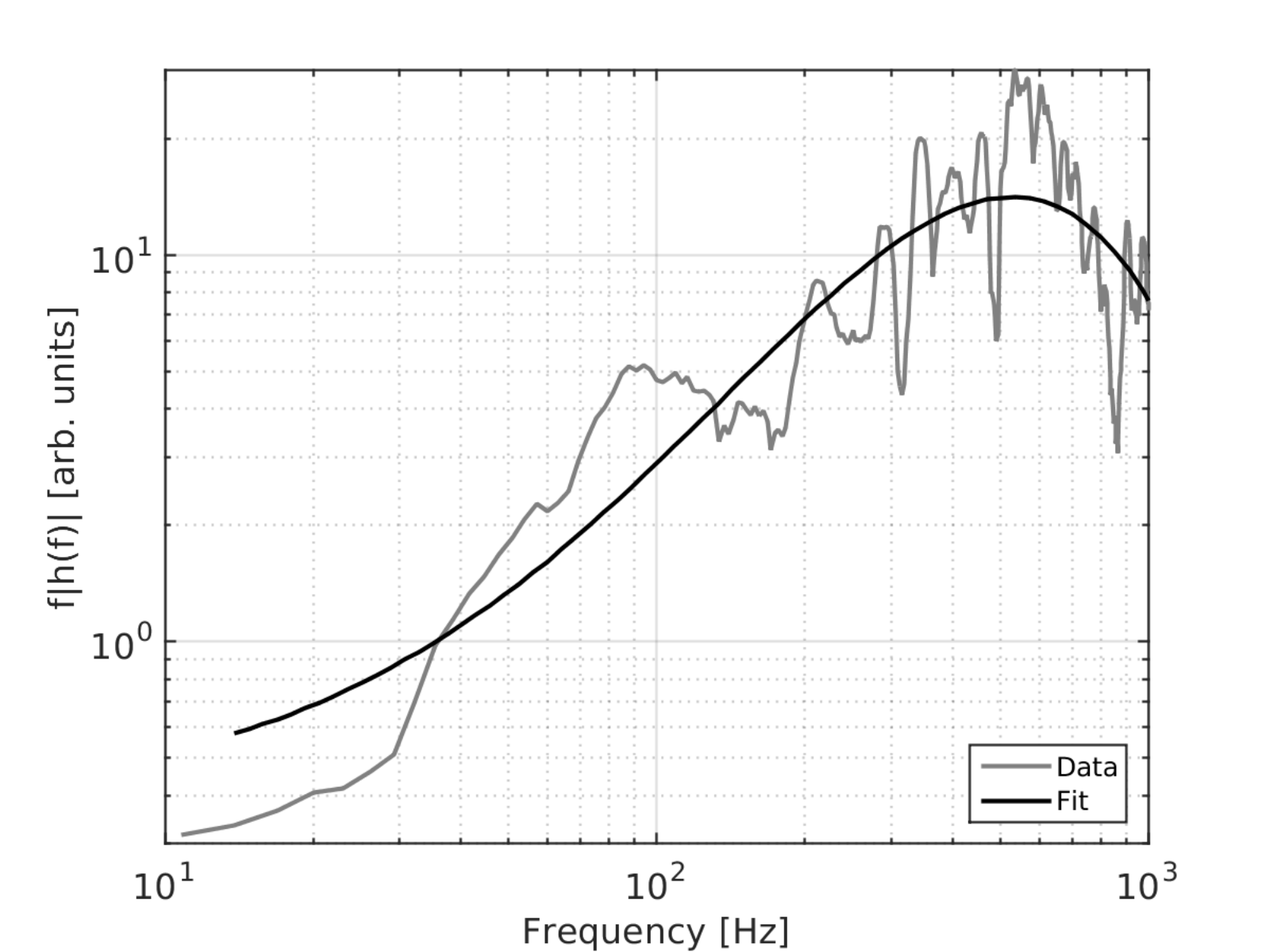}
  \caption{Gravitational wave signal from an equatorial observation computed by M{\"u}ller et al. in the \texttt{u8.1} simulation~\cite{muller2013}. The original $f|\tilde{h}(f)|$ data are plotted in gray, and the fit to these data (with $a = 75$ Hz and $b = 201$ Hz) by our $f|\tilde{h}(f)|$ model (Eq.~\ref{funcform}) is shown in black.
      }
  \label{no_lowpeak_mul}
\end{figure}

Finally, we use spectra from simulations performed by M{\"u}ller et al.~\cite{muller2013}. These simulations were carried out using the \texttt{Vertex-CoCoNuT}~\cite{muller2010} or \texttt{Vertex-Prometheus}~\cite{Buras_neutrino_transport, rampp_hydrodynamics} codes.
These are similar neutrino hydrodynamics codes, but \texttt{Vertex-CoCoNuT} uses general relativity while \texttt{Vertex-Prometheus} is pseudo-Newtonian.
The simulations model core-collapse supernovae in two dimensions, and include neutrino transport using the ray-by-ray plus approximation \cite{Buras_neutrino_transport} and a variable Eddington factor technique via \texttt{Vertex}, the neutrino transport module~\cite{rampp_hydrodynamics}.
Six zero-age main sequence progenitor masses were assumed, from $8.1 {\rm \; M_{\odot}}$ to $27 {\rm \; M_{\odot}}$.
All models use the Lattimer-Swesty EOS~\cite{lattimer-swesty}, three with $K = 220$ MeV and three with $K = 180$ MeV~\cite{muller2013}.
Figure \ref{no_lowpeak_mul} shows the strain spectrum from the $8.1 {\rm \; M_{\odot}}$ progenitor evolved using GR and an EOS with $K = 180$ MeV, as well as our fit to this data.

Studies of these simulated GW spectra suggest that the following parameter ranges encapsulate most of the variation in the GW strain spectra:
\begin{itemize}
  \item $5 {\rm \; Hz} < a < 150 {\rm \; Hz}$
  \item $10 {\rm \; Hz} < b < 400 {\rm \; Hz}$.
\end{itemize}
The fits are qualitative, but generally fall within a factor of two of the simulation data.
The scaling parameter $\xi$ is not restricted by these fits since it is related to the overall amplitude of the SGWB energy density spectrum rather than to the individual source strain spectrum morphology.

\subsection{Low-Frequency Model}

The spectra shown in Figures \ref{no_lowpeak_ott}-\ref{no_lowpeak_mul} were fitted well with the functional form of Eq.~\ref{funcform}. However, we observed that in some simulations the emitted GW spectra feature two peaks, one at high frequencies, which is well-modeled by Eq.~\ref{funcform}, and another at $\approx$ 60--120 Hz.
Of the 64 numerical core collapse simulations we considered, over 80\% exhibited signs of the low-frequency peak, some more prominently than others.

The precise physical process that produces this low-frequency peak in the GW spectrum is unclear, but several papers have made note of it and attempted to identify its source. M{\"u}ller et al. \cite{muller2013} and Yakunin et al. \cite{yakunin} consider the deflection of infalling matter through the shock as a major contributor to this peak, but ultimately rule it out. M{\"u}ller et al. \cite{muller2013} go on to indicate that a quasi-periodic signal around $100$ Hz arises due to prompt post-shock convection during the first tens of milliseconds following core bounce. However, they rule out convection as the direct source of this signal because the entropy and lepton gradients which drive the convection quickly disperse, while the GW emission lasts for tens of milliseconds; instead, they speculate that the convection leads to the development of acoustic waves which produce a GW signal due to the resulting mass motions. Finally, they note that this signal always peaks near $100$ Hz in their simulations; there appears to be no significant dependence on the properties of the progenitor.

Ott et al. \cite{ott2013} found that early GW emission due to prompt convection sets in at $\approx 10$ ms after core bounce and lasts for about $30$ ms. They also note that the GW signal produced by prompt convection is particularly sensitive to the perturbations which drive the convection; this is compatible with the statement from M{\"u}ller et al. \cite{muller2013} that the signal from prompt convection may depend on the width of the layer in which the convection occurs.

Kuroda et al. \cite{kuroda2014} found that the speed of core rotation during the collapse process may impact the frequency of the signal from prompt convection. Simulations in which the core was rapidly rotating led to the emission of spiral waves; the acoustic waves which produce the low-frequency GW signal were Doppler shifted by emission on top of the spiral waves, leading to a low-frequency peak closer to 200 Hz. They also note that accurate neutrino transport is a necessary component of these simulations in order to have a reliable prediction of the GW signal; this agrees with similar findings from M{\"u}ller et al. \cite{muller2013} which found significantly reduced GW emission at low frequencies when simplified neutrino rates were used.

Finally, recent 3D simulations by Andresen et al.~\cite{andresen2016} indicate the presence of an additional low-frequency component of the GW signal, which appears to be related to SASI.
They speculate that the convection-related GW signal in the 100--200 Hz range may be actually driven by the coupling of SASI dynamics to the stellar interior.
They also note that the signals from their 3D simulations are considerably different from those of 2D simulations, and amplitudes are significantly lower, especially at higher frequencies.

To summarize, there is significant evidence that the low-frequency peak is a realistic feature of the GW signal produced during core-collapse; it is present in many different simulations and there is a general agreement that the signal is related to prompt postbounce convection or SASI.
Although it seems to vary for the different simulations, the peak in the GW frequency spectrum tends to occur between $\approx$60--120 Hz, or even as high as 200 Hz for the rapidly-rotating models of \cite{kuroda2014}.
We also note that not all of the simulations included in our study exhibited this peak; however, it seems that the more modern simulations, which feature general relativistic hydrodynamics and sophisticated neutrino transport, generally include this feature \cite{tannerThesis}.

\begin{figure}
  \includegraphics[width=3.2in]{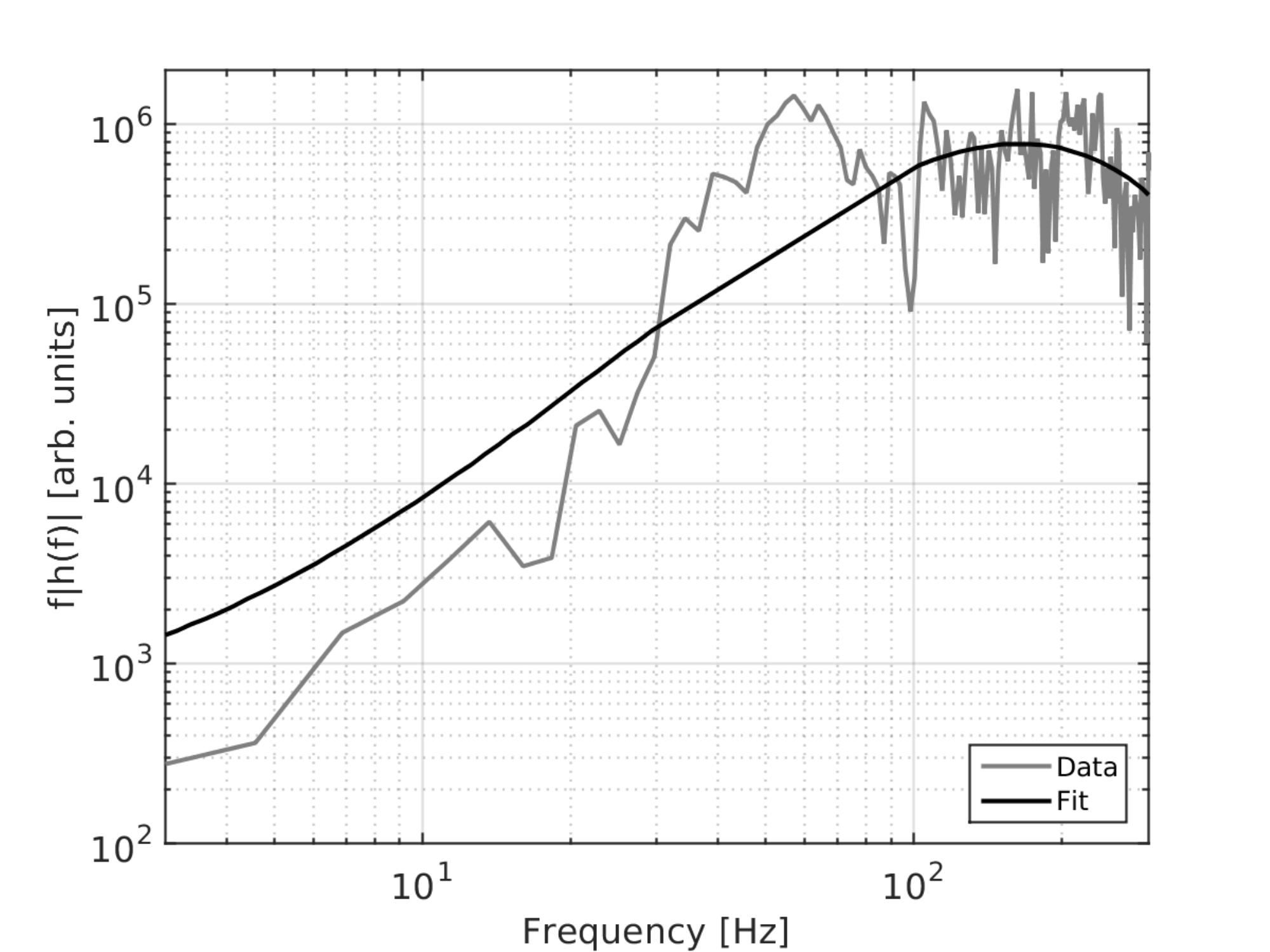}
  \caption{Gravitational wave signal from the Ott et al. \texttt{s27f$_\text{heat}$1.00} simulation~\cite{ott2013}. In particular, this signal is from a polar observation of a plus-polarized gravitational wave. The original $f|\tilde{h}(f)|$ data is plotted in gray, and the fit to this data (with $a = 5$ Hz and $b = 56$ Hz) to our $f|\tilde{h}(f)|$ model in Eq.~\ref{lowpeak} is shown in black. Note the low frequency structure at $\approx 30$--$100$ Hz, which deviates from the functional form of Eq.~\ref{lowpeak}, necessitating a modified functional form to describe this spectrum accurately.}
  \label{lowpeak}
\end{figure}

\begin{figure}
  \includegraphics[width=3.2in]{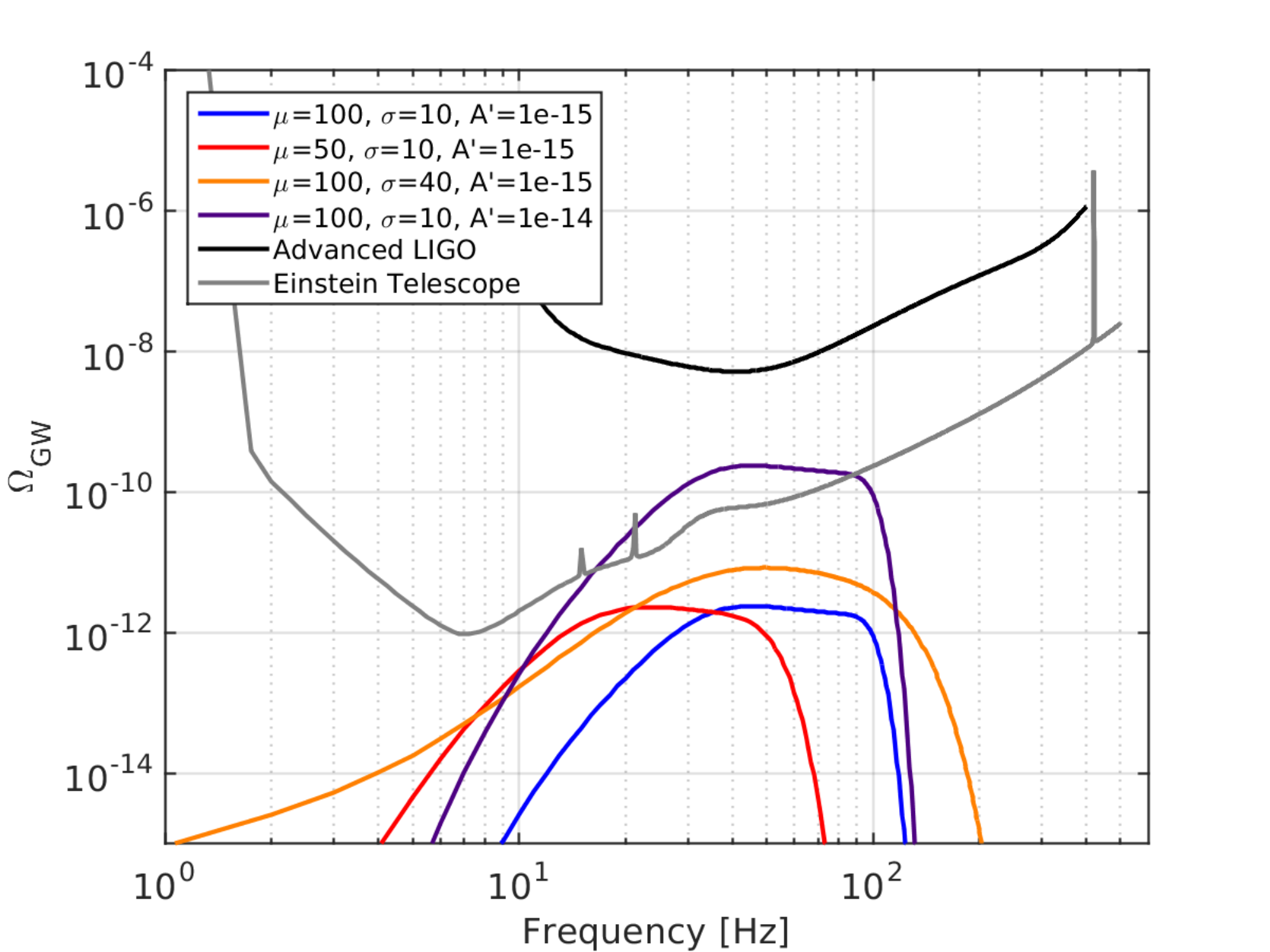}
  \caption{$\Omega_\text{GW}(f)$ for various parameter choices for the low-frequency model of an SGWB produced by stellar core-collapse. $\mu$ and $\sigma$ are in units of Hz, while $A'$ is in units of m/kg$^{1/2}$. In Eq.~\ref{eq:low}, $\sigma$ defines the width of the Gaussian peak and has no effect on the peak amplitude of the strain spectrum; however, when it is integrated over redshift to calculate $\Omega_\text{GW}$, a wider peak results in an increased GW energy density. A larger value of $\sigma$ also allows for some amount of GW energy density at frequencies above $\mu$, while low values of $\sigma$ result in a sharp cutoff at $f = \mu$. Finally, $\Omega_\text{GW}(f)$ goes as $A'^2$, so this is simply a scaling parameter~\cite{tannerThesis}.}
  \label{low_model}
\end{figure}

Figure \ref{lowpeak} shows that the functional form of Eq.~\ref{funcform} is not sufficient to capture the low-frequency behavior of the simulated spectrum. This issue is particularly important because the terrestrial GW detectors have optimal sensitivities below 100 Hz, as shown in Figure \ref{high_frequency}. We model the low-frequency behavior with a Gaussian function in the strain spectrum governed by three parameters. The mean, denoted $\mu$, defines the center of the Gaussian peak, the standard deviation, denoted $\sigma$, defines the width of the peak, and $A$ denotes the amplitude of this peak:
\begin{eqnarray} \label{low_freq_strain}
f_e|\tilde{h}(f_e)|_\text{low} = \frac{A}{D} \exp{\bigg(- \frac { (f_e - \mu)^2}{2 \sigma^2}\bigg)}
\end{eqnarray}
where $A$ is scaled by $D$, the distance from the star at which the signal is observed (assumed to be sufficiently small for redshifting effects to be neglected).
The corresponding GW energy spectrum is given by:
\begin{eqnarray} \label{eq:low}
\frac{dE}{df_e}(f_e)_\text{low} = \frac{\pi^2 c^3 A^2}{G} \exp{\bigg(- \frac { (f_e - \mu)^2}{2 \sigma^2}\bigg)}
\end{eqnarray}
which gives GW energy density:
\begin{eqnarray}
\Omega_\text{GW} (f) = \frac{8 \pi^3 c}{3H_0^2} f A'^2 \int dz \frac{R_* (z)}{(1+z)E(\Omega_\text{m},\Omega_{\Lambda},z)}  \nonumber \\\exp{\left(- \frac{ (f(1+z) - \mu)^2}{\sigma^2}\right)}
\end{eqnarray}
This equation combines the two scaling parameters $A$ and $\lambda_\text{CC}$ into $A' = A \lambda_\text{CC}^{1/2}$, with units of m/kg$^{1/2}$ \cite{tannerThesis}.
Figure \ref{low_model} shows this model for several choices of parameters.

\subsection{Combined High and Low-Frequency Model}

We also consider a combined case, which includes both the low-frequency peak and the higher-frequency functional form for the core-collapse spectrum. Following \cite{tannerThesis}, this results in an energy spectrum of
\begin{align}\label{combined}
  \begin{split}
  \frac{dE}{df_e}(f_e)_\text{comb} &= \frac{\xi}{\lambda_\text{CC}}\Bigg[ \pi A' \sqrt{ \frac{c^3}{\xi G}} e^{- \frac{ (f_e - \mu)^2}{2 \sigma^2}} + \\
  &\phantom{=} \left(1 + \frac{f_e}{a} \right)^3 e^{-f_e/b} \Bigg]^2.
  \end{split}
\end{align}
\begin{eqnarray}
  \Omega_\text{GW}(f) & = &\frac{8 \pi G}{3H_0^3 c^2} f \frac{\xi}{\lambda_\text{CC}} \int dz \frac{R_* (z)}{(1+z)E(\Omega_\text{m},\Omega_{\Lambda},z)} \nonumber \\
  & \times &  \Bigg[ \pi A' \sqrt{ \frac{c^3}{\xi G}} e^{-\frac { (f(1+z) - \mu)^2}{2 \sigma^2}} \nonumber \\
  & + & \left(1 + \frac{f(1+z)}{a} \right)^3 e^{-f(1+z)/b} \Bigg]^2
\end{eqnarray}

\section{Results}\label{sec:results}

\subsection{High Frequency Model}

As discussed above, the free parameters of the high-frequency SGWB model are $\xi$, $a$, and $b$. The $\xi$ parameter is simply a scaling factor related to the total energy emitted in GWs, and the $a$ and $b$ parameters effectively determine the shape of the spectrum: its rise, peak position, and drop-off at high frequencies.
We scan the parameter space $(a,b,\xi)$ of this model, restricting $5 {\rm \; Hz} < a < 150 {\rm \; Hz}$ and $10 {\rm \; Hz} < b < 400 {\rm \; Hz}$ as discussed in the previous section.
The range of $\xi$ is tuned so as to optimally probe the accessibility of the model to each detector pair.
For Advanced LIGO, we use a range of $\xi = 10^6 \text{--} 10^{13}  {\rm \; m^2/s}$, and for ET we use $\xi = 10^2 \text{--} 10^9 {\rm \; m^2/s}$~\cite{tannerThesis}.
For each point in this parameter space, we compute the spectrum $\Omega_{\rm GW}(f)$ and compare it to the sensitivities of the Advanced LIGO and ET detectors, computing the likelihood function:
\begin{eqnarray}
  \mathcal{L} \propto \displaystyle{\prod_i} \exp\left[-\frac{(Y_i - \Omega_{\text{GW},i}(\xi,a,b))^2}{2 \sigma_i^2}\right],
\label{likelihood_eq}
\end{eqnarray}
where the index $i$ runs over frequency bins, $Y_i$ is the expected measurement of the GW energy density in the bin $i$, $\sigma_i$ is the corresponding measurement error, and $\Omega_{{\rm GW},i}(\xi,a,b)$ is the modelled energy density in the bin $i$ for the given free parameters $\xi$, $a$, and $b$. For projecting the future experimental sensitivities we set $Y_i=0$. In order to determine the accessibility of this three-dimensional parameter space to future detectors, we marginalize (integrate) the likelihood function over one of the parameters, and then compute the expected sensitivity contours at 95\% confidence in the plane of the remaining two parameters.

\begin{figure*}[!t]
  \centering
  \includegraphics[width=3.2in]{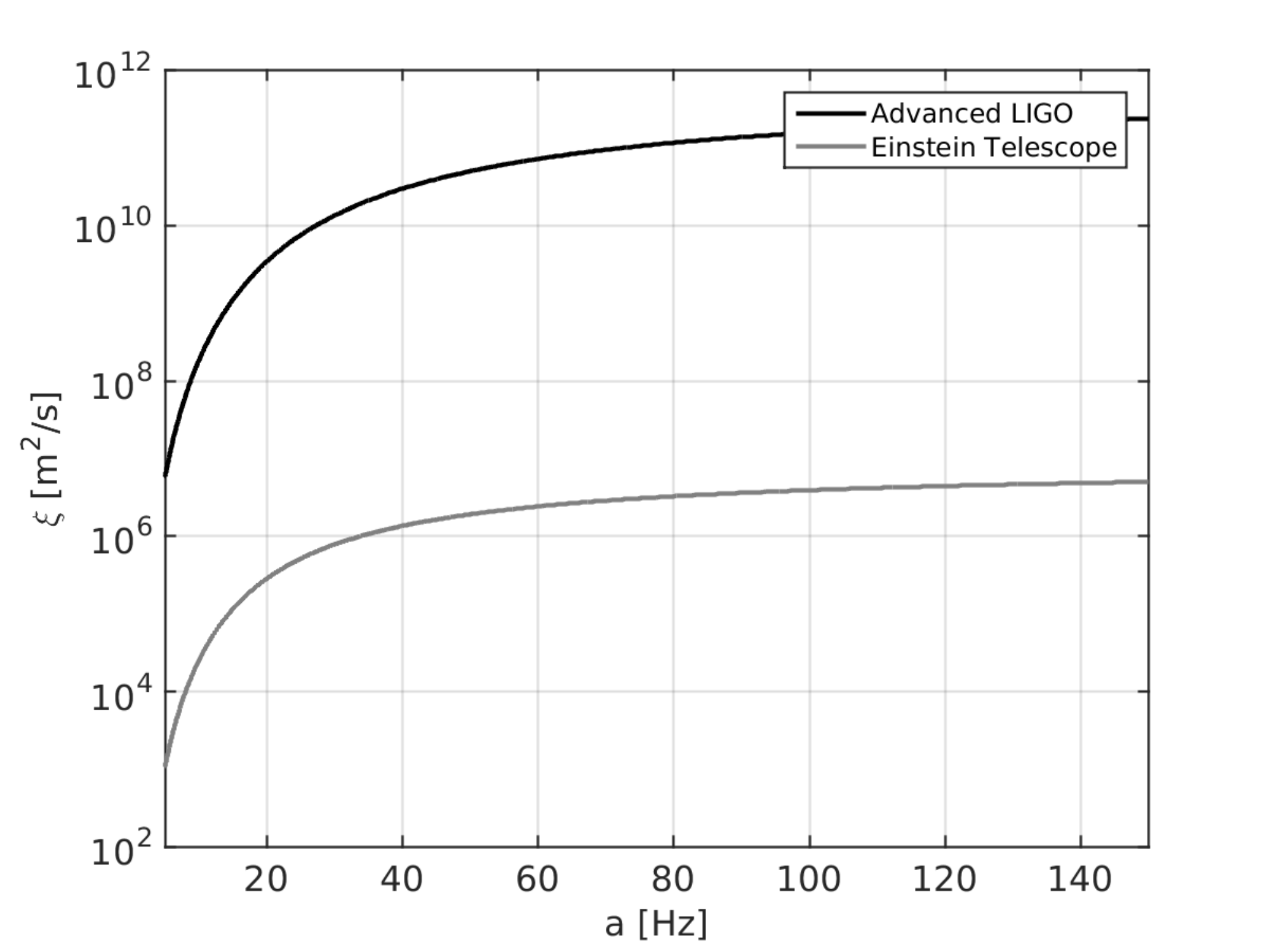} \quad
  \includegraphics[width=3.2in]{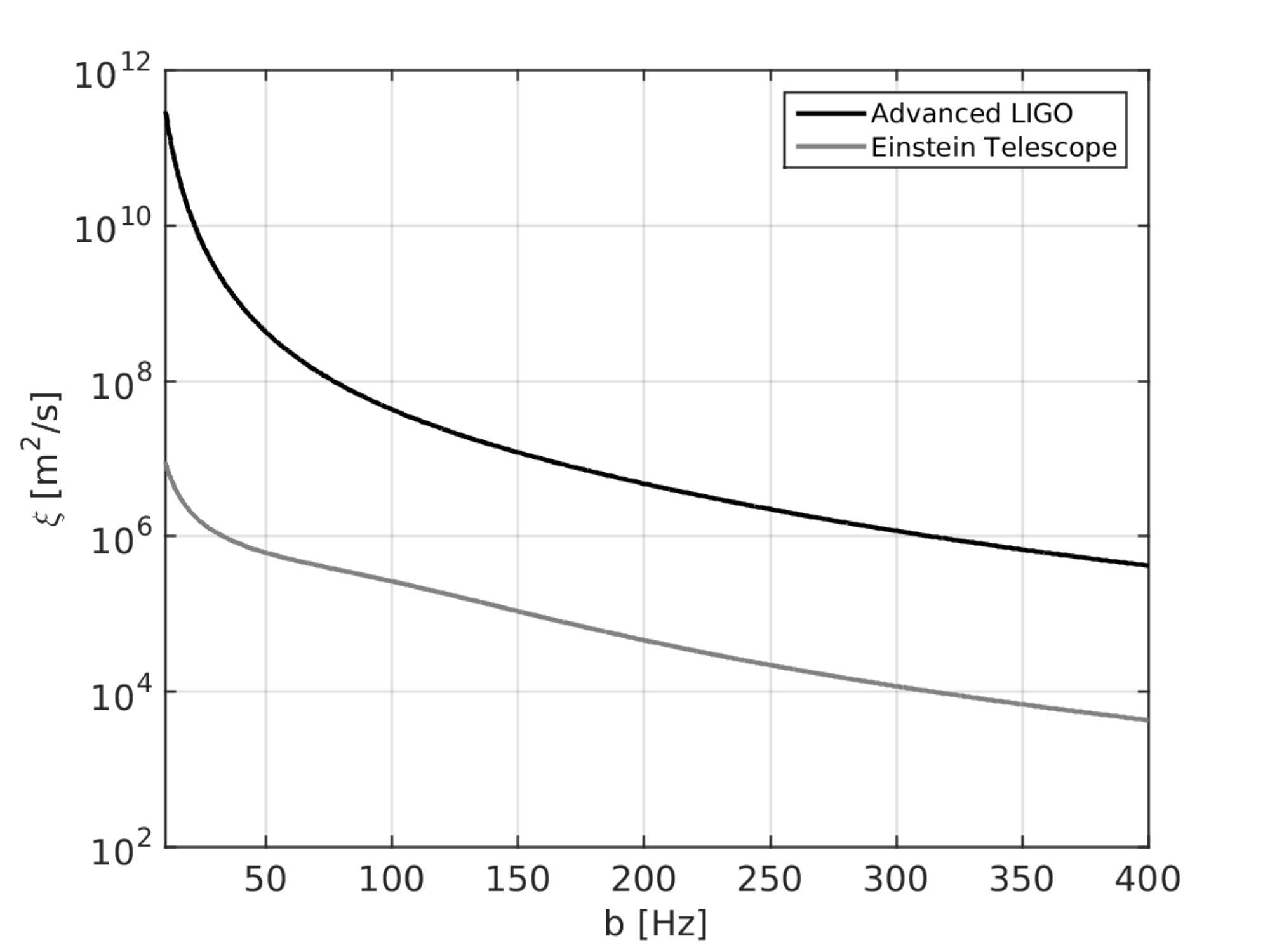} \\
  \includegraphics[width=3.2in]{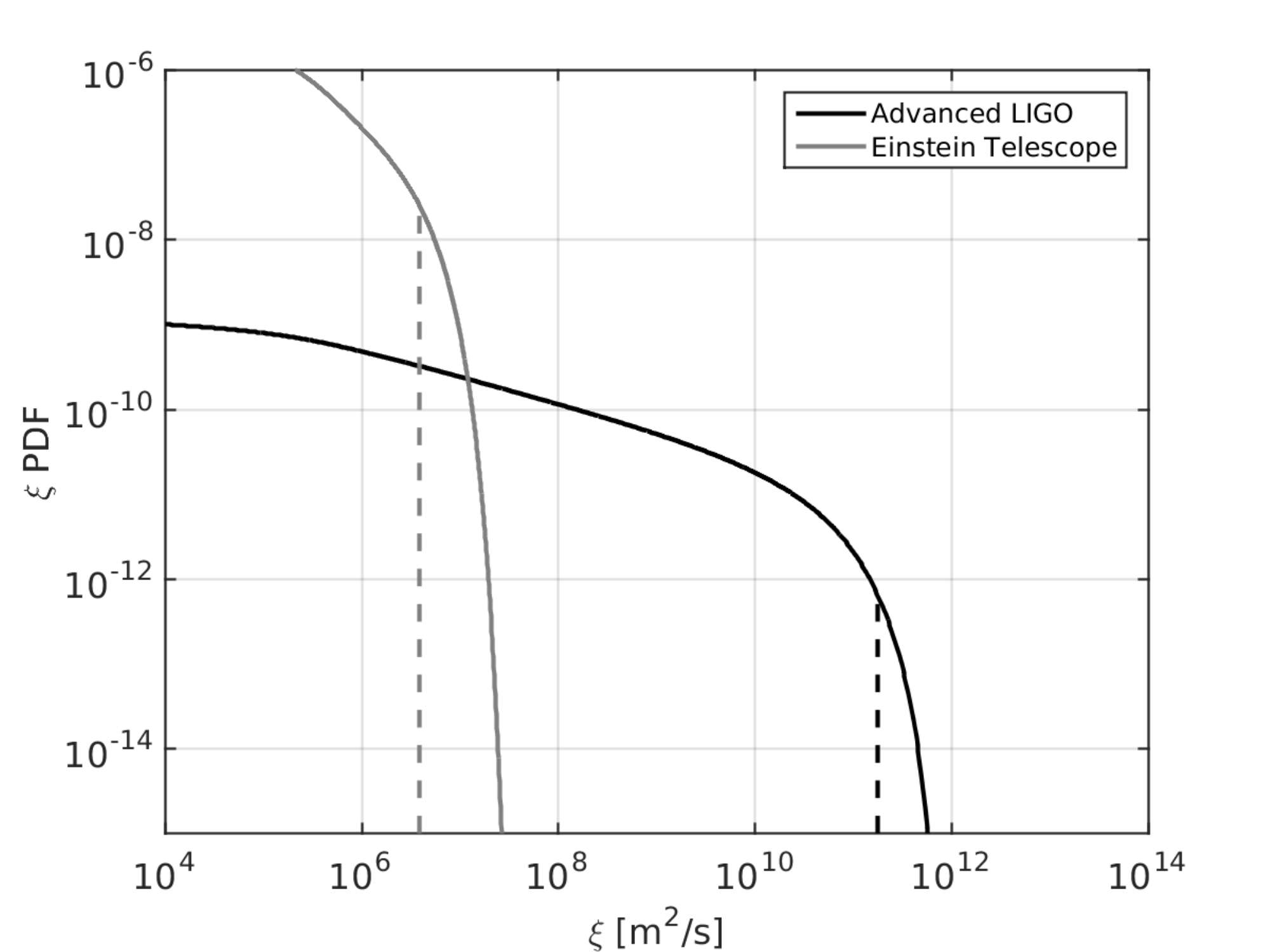} \quad
  \includegraphics[width=3.2in]{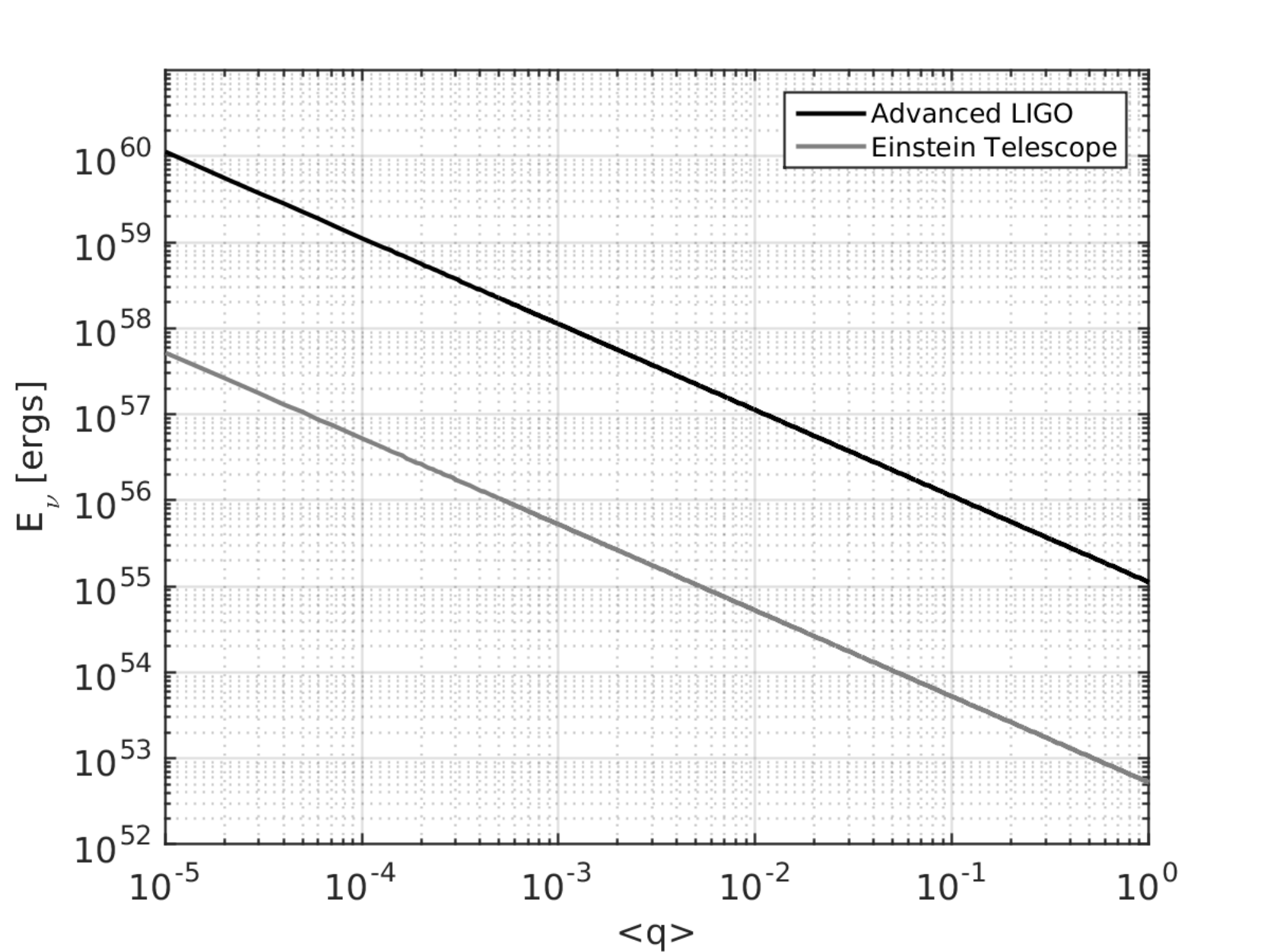}
  \caption{Top-left: 95\% confidence expected sensitivity contours for Advanced LIGO and Einstein Telescope in the $\xi$-$a$ plane, after marginalizing over the $b$ parameter. Top-right: 95\% confidence expected sensitivity contours for Advanced LIGO and Einstein Telescope in the $\xi$-$b$ plane, after marginalizing over the $a$ parameter. Bottom-left: posterior distribution of $\xi$ after marginalizing over all other parameters; expected sensitivities at 95\% confidence are shown as dashed vertical lines. Bottom-right: sensitivity in $\xi$ is translated into sensitivity in the $E_{\nu}$-$\langle q \rangle$ plane, assuming $\lambda_\text{CC} \approx 0.01$ M$_{\odot}^{-1}$. More details are provided in the text~\cite{tannerThesis}.}
  \label{model_contours}
\end{figure*}

Figure \ref{model_contours} shows the resulting 95\% expected sensivitity contours for both Advanced LIGO and Einstein Telescope in the $\xi$-$b$ and $\xi$-$a$ planes.
%Curves for both Advanced LIGO and Einstein Telescope, and for both the GRB-based and the luminosity-based star formation rates are shown.
The curves in the $\xi$-$b$ plane are decreasing with $b$, which is a consequence of the fact that increasing $b$ cuts off the spectrum at higher frequencies, hence making the model more detectable by the GW detectors. Similarly, lowering the value of $a$ allows the spectrum to rise at lower frequencies (c.f. Figure \ref{high_frequency}), making the model more accessible to GW detectors and causing the increasing trend (with $a$) of the curves in the $\xi$-$a$ plane.
%The two star formation rates yield nearly identical predictions, which is the consequence of the fact that the dominant contribution to $\Omega_{\rm GW}$ comes from redshifts region of 1-2, in which the two models of star formation rate agree well
It is also evident that the Einstein Telescope will provide a substantially better probe of this model than the second-generation detectors.

Marginalizing over the $a$ and $b$ parameters yields the posterior distribution for the amplitude parameter $\xi$.
The 95\% expected sensitivity on $\xi$ is $1.8 \times 10^{11}$ m$^2$/s and $3.8 \times 10^6$ m$^2$/s for Advanced LIGO and Einstein Telescope, respectively.
For the representative value of the mass fraction parameter $\lambda_\text{CC} \approx 0.01$ M$_{\odot}^{-1}$, these sensitivities can be translated into sensitivity curves in the $E_{\nu}$-$\langle q \rangle$ plane, also shown in Figure \ref{model_contours} (bottom).
For a neutrino asymmetry of $\langle q \rangle = 0.0045$ (following \cite{firststars, buonanno}), we see that Advanced LIGO would require the total neutrino energy to be approximately $2.5 \times 10^{57}$ ergs for a detectable SGWB, which is rather high.
Einstein Telescope would require $E_{\nu} \approx 1.2 \times 10^{55}$ ergs, which is still about two orders of magnitude above expected neutrino energies from the literature ($\approx 3 \times 10^{53}$ ergs \cite{firststars, buonanno}).
It is important to note, however, that these sensitivity numbers are obtained after integrating over a large parameter space.
For example, it is evident from the top two panels of Figure \ref{model_contours} that there is a substantial part of the parameter space with $\xi <  3.8 \times 10^6$ m$^2$/s that would still be accessible to Einstein Telescope, hence potentially including the required neutrino energy level.

\subsection{Low Frequency Model}

The ranges for the low frequency model parameters are based on the previously described work with the individual simulations. For most of the simulations, the low frequency peak occurs between 60--120 Hz; however, there are some cases described in the literature where this peak may occur as high as 200 Hz due to rapid rotation during the collapse process \cite{kuroda2014}. Thus, we have used a liberally defined range on $\mu$, 30--200 Hz. Most of the waveforms that we studied from \cite{ott2013} had relatively narrow low-frequency peaks, with typical widths between 40--60 Hz. The low-frequency peaks in the waveforms from \cite{abd2014} tended to be wider, with some as broad as $140$ Hz. We assume the range of $\sigma$ to be 10--80 Hz. Because $A'$ is a scaling parameter like $\xi$, its scanned range was tuned to each detector pair: $A' = 10^{-15}$--$10^{-12}$ m/kg$^{1/2}$ for Advanced LIGO and $A' = 10^{-17}$--$10^{-13}$ m/kg$^{1/2}$ for Einstein Telescope.

\begin{figure*}[!t]
    \centering
    \includegraphics[width=3.2in]{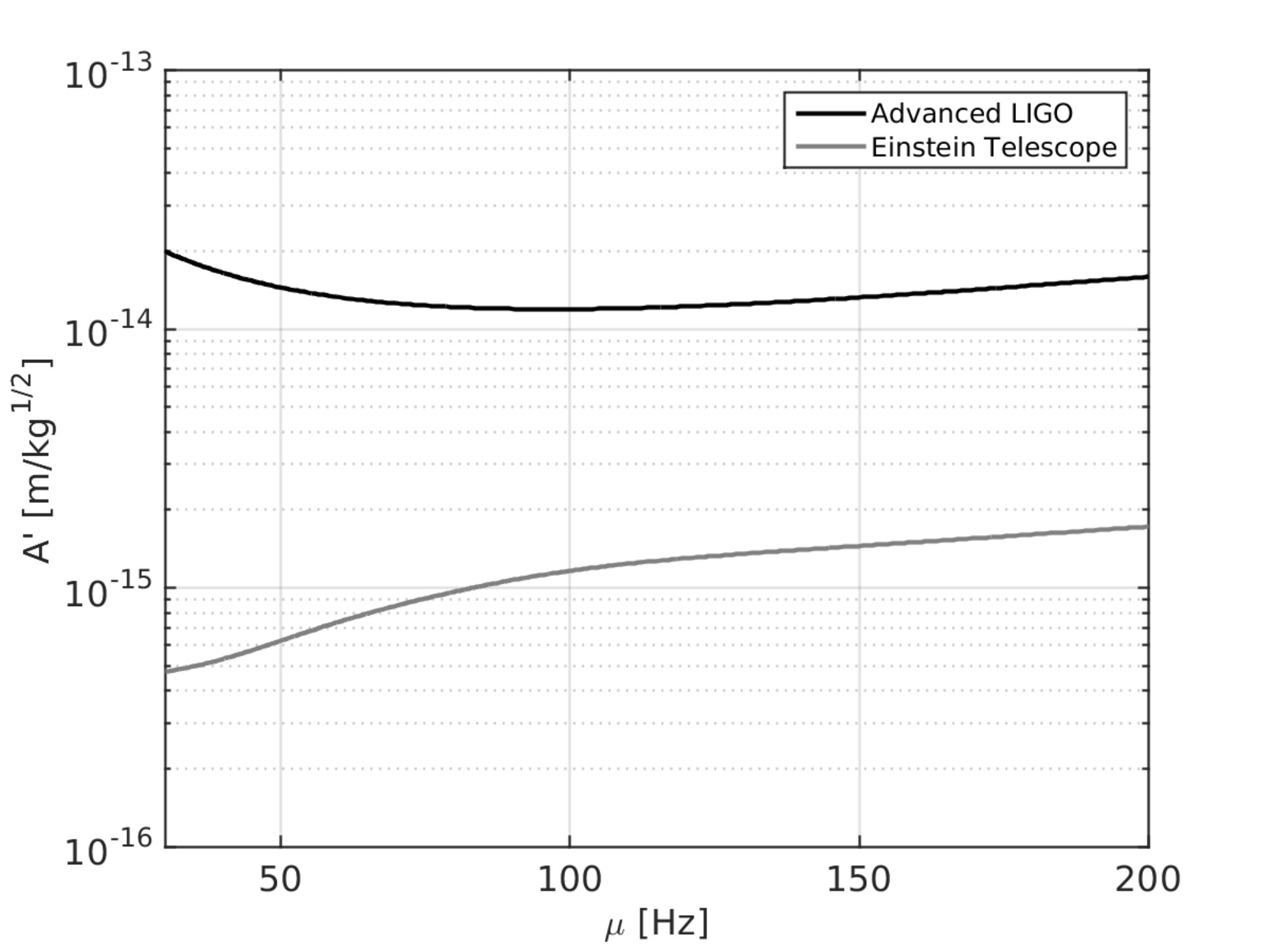} \quad
    \includegraphics[width=3.2in]{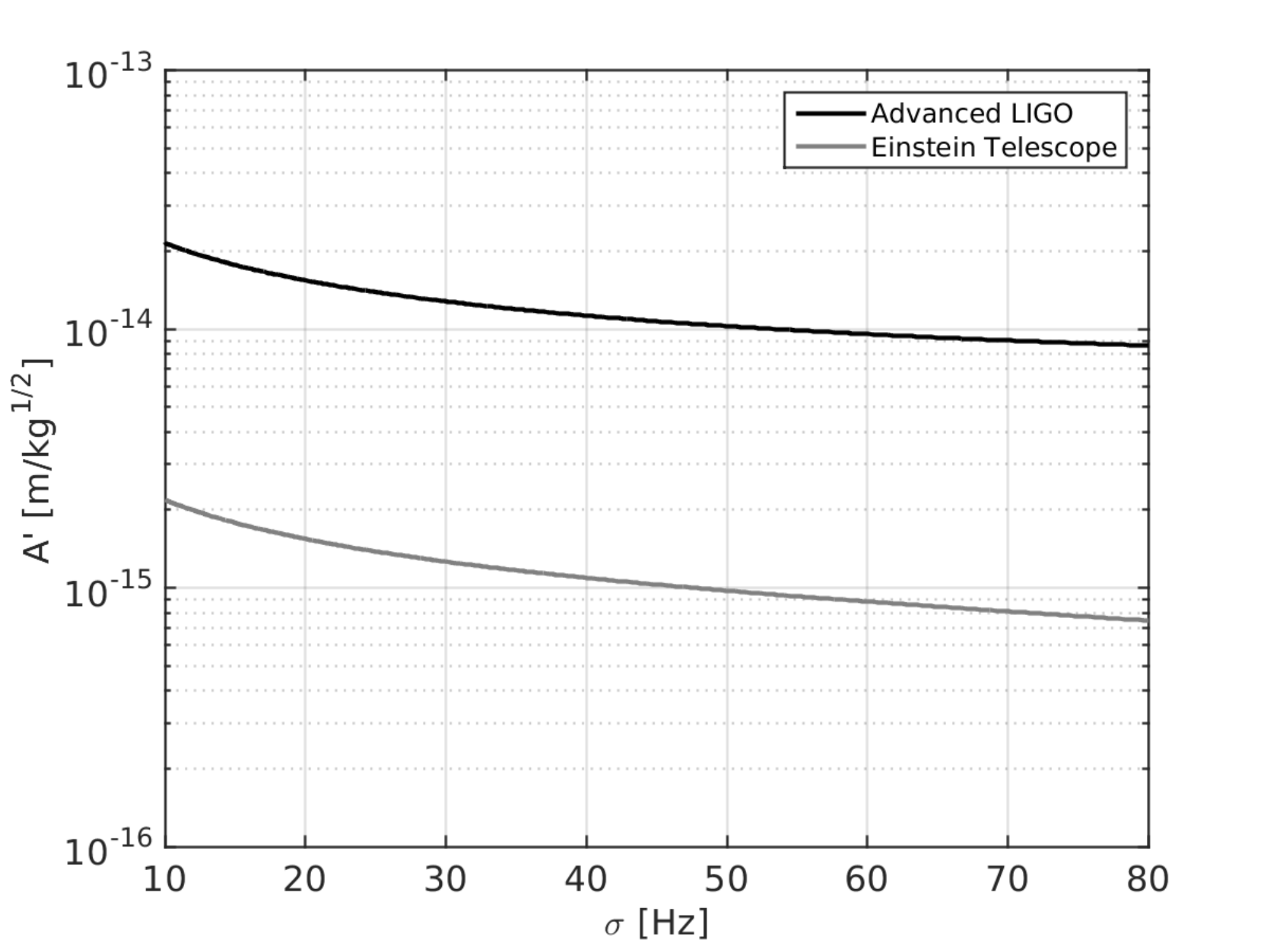} \\
    \includegraphics[width=3.2in]{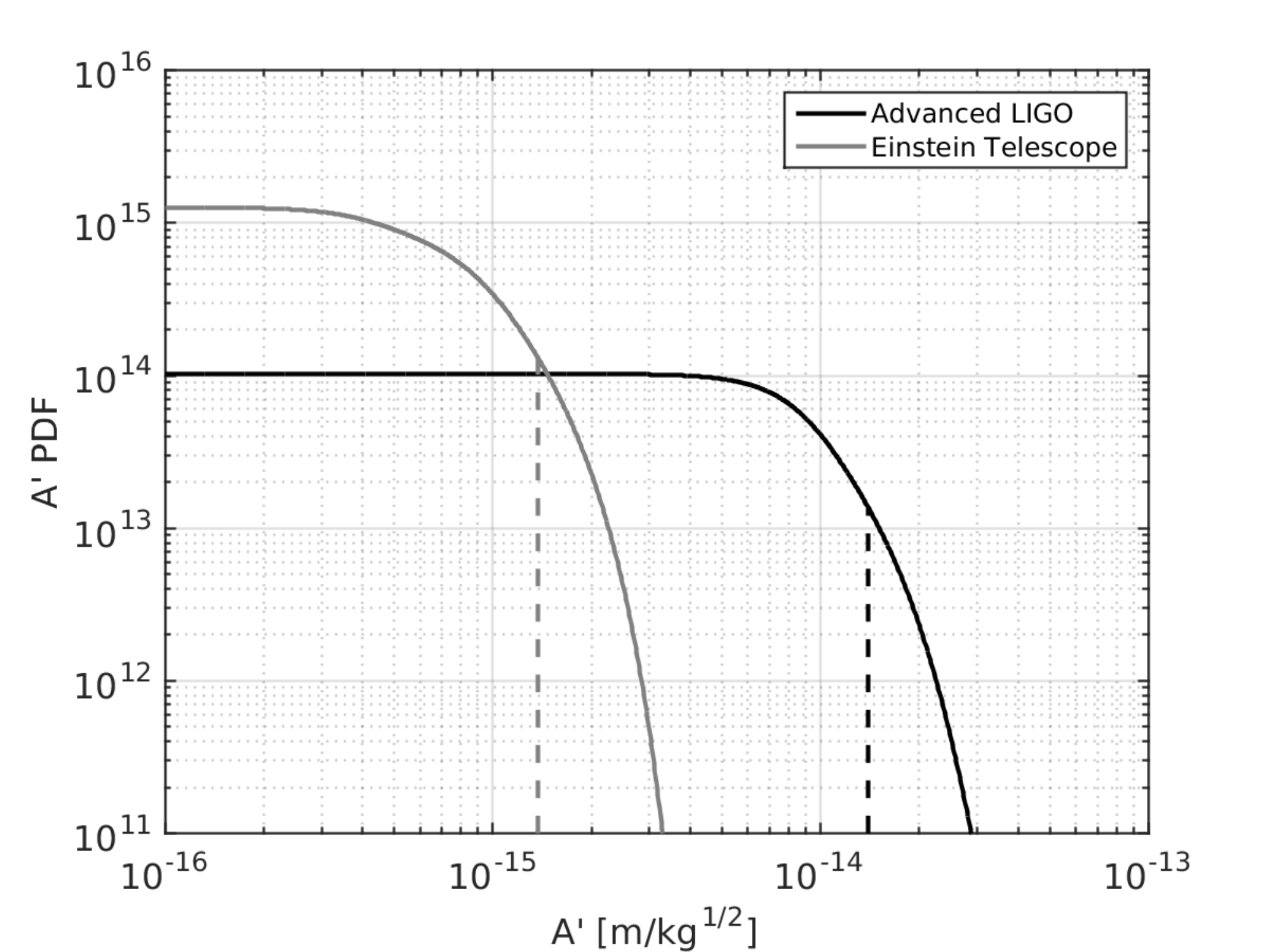} \quad
    \includegraphics[width=3.2in]{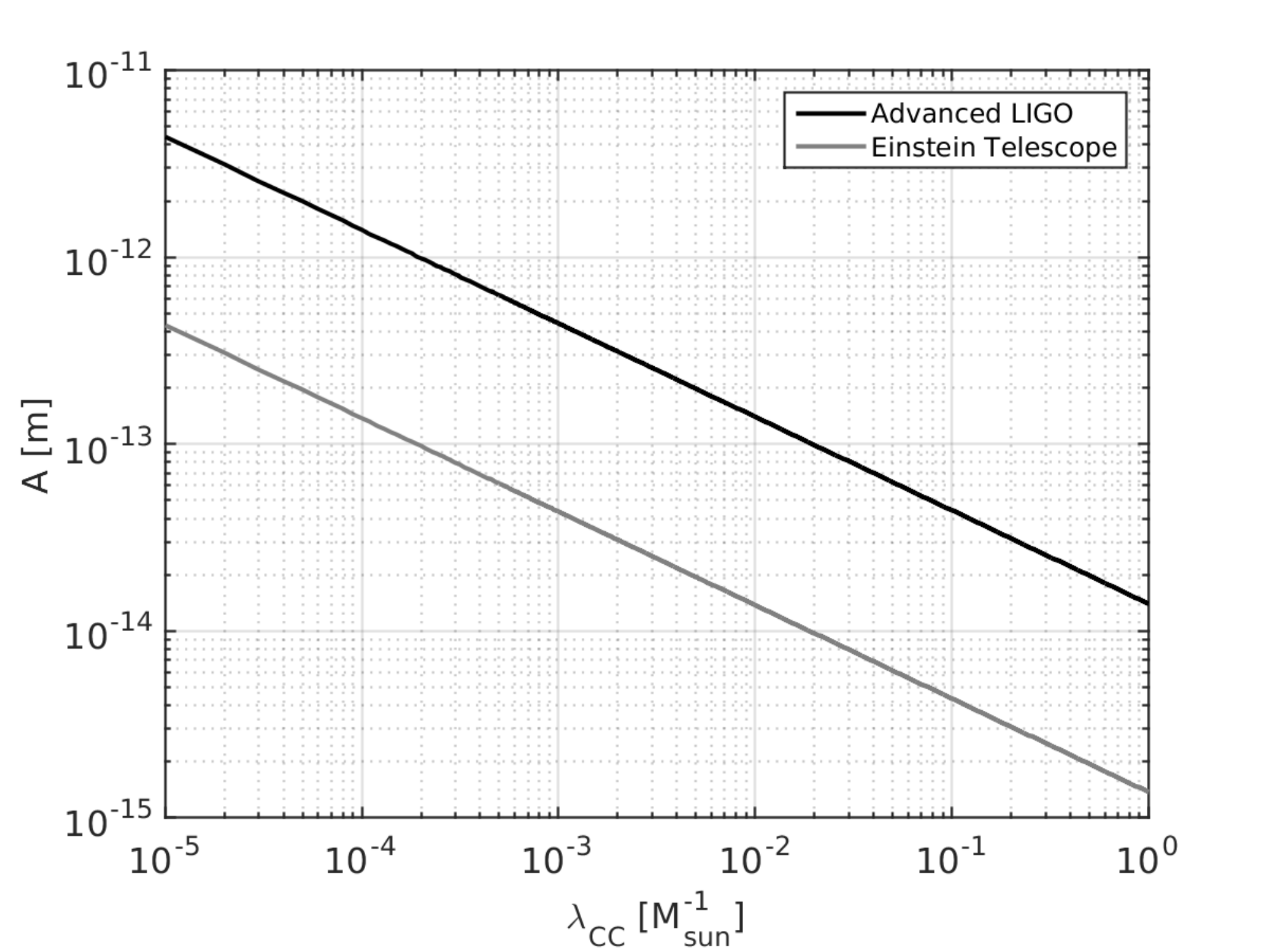}
    \caption{Top-left: 95\% confidence expected sensitivity contours for Advanced LIGO and Einstein Telescope in the $A'$-$\mu$ plane, after marginalizing over the $\sigma$ parameter. Top-right: 95\% confidence expected sensitivity contours for Advanced LIGO and Einstein Telescope in the $A'$-$\sigma$ plane, after marginalizing over $\mu$. Bottom-left: posterior distribution of $A'$ after marginalizing over all other parameters; expected sensitivities at 95\% confidence are shown as dashed vertical lines. Bottom-right: sensitivity in $A'$ is translated into sensitivity in the $A$-$\lambda_\text{CC}$ plane. More details are provided in the text~\cite{tannerThesis}.}
  \label{low_model_amplitude}
\end{figure*}

We have performed a scan of the parameter space using an identical method to that described in the previous section. Figure \ref{low_model_amplitude} shows 95\% sensitivities in two-dimensional parameter spaces where we have marginalized over the third model parameter. Regions of parameter space above each of the curves should be detectable by Advanced LIGO (black) or the Einstein Telescope (gray). In the $A'$-$\mu$ plane, the Advanced LIGO curve reaches a minimum near $\mu = 80$ Hz, since this is where the stochastic search achieves its best sensitivity with this detector (see Figure \ref{high_frequency}). However, the Einstein Telescope is most sensitive below 10 Hz, which is not included in our parameter space for $\mu$. As a result, this contour decreases with decreasing $\mu$. For both detectors, the contours in the $A'$-$\sigma$ parameter space decrease as $\sigma$ increases. This is because $\sigma$ acts similarly to a scaling parameter for $\Omega_\text{GW}(f)$, as previously discussed.

We also marginalize over the $\mu$ and $\sigma$ parameters to estimate an expected sensitivity on $A'$ at 95\% confidence.
These results indicate that Advanced LIGO should be sensitive to cases with $A' > 1.4 \times 10^{-14} {\rm \; m/kg^{1/2}}$ and Einstein Telescope should be sensitive to cases with $A' > 1.4 \times 10^{-15}{\rm \; m/kg^{1/2}}$.
In the bottom panel of Figure \ref{low_model_amplitude}, we have used these expected sensitivities on $A'$ to calculate contours in the $A$-$\lambda_\text{CC}$ parameter space.
Taking $\lambda_\text{CC}$ to be 0.01 M$_\odot^{-1}$ gives sensitivities of $A>$ 197 m for Advanced LIGO and  $A>$ 19.5 m for the Einstein Telescope.

For comparison, we examine the \texttt{s27f$_\text{heat}$1.00} simulation of \cite{ott2013} and the \texttt{A1O05.5} simulation of \cite{abd2014}.
For each simulation, we compute the strain spectrum $f|\tilde{h}(f)|$ and fit the functional form from Eq.~\ref{high_freq_strain} to the simulation (not including the frequencies where the low-frequency structure is prevalent).
Next, we subtract this fit from the data and then fit the Gaussian functional form given in Eq.~\ref{low_freq_strain} to the low-frequency residual.
The best fits for this functional form correspond to values of $A = 0.18$ m for the \texttt{s27f$_\text{heat}$1.00} simulation and $A = 0.85$ m for the \texttt{A1O05.5} simulation.
These values are approximately 3 and 2 orders of magnitude below the expected Advanced LIGO sensitivity to $A$, respectively, and 2 and 1 orders of magnitude below the expected Einstein Telescope sensitivity, implying that these experiments are unlikely to detect this signal.
%Increasing the assumed value for $\lambda_\text{CC}$ from $0.01 {\rm \; M_{\odot}^{-1}}$ to $1 {\rm \; M_{\odot}^{-1}}$ would improve the resulting sensitivities by a factor of 10; however, it seems unrealistic to expect one core-collapse per solar mass of star-forming material. Even in this extreme case, the expected sensitivities would still be larger than the values of $A$ given by the fits to the simulations. Thus, we do not expect Advanced LIGO or the Einstein Telescope to be sensitive to these types of signals unless our assumptions about the rate of core-collapse supernovae are relatively inaccurate. \cite{tannerThesis}

\subsection{Combined Case}

We have also performed a parameter estimation study for the combined model, which is the sum of the low-frequency peak and the higher frequency model, as shown in Eq.~\ref{combined}. In this case, the model has six parameters, and a brute force exploration of the model parameter space is computationally infeasible. In order to study this model, we employ the \texttt{MultiNest} algorithm~\cite{multinest1, multinest2, multinest3}. The same parameter ranges given in the previous two sections are used in this combined study.

\begin{figure*}[!t]
	\begin{tabular}{cc}
	\includegraphics[width=3.2in,height=2.5in]{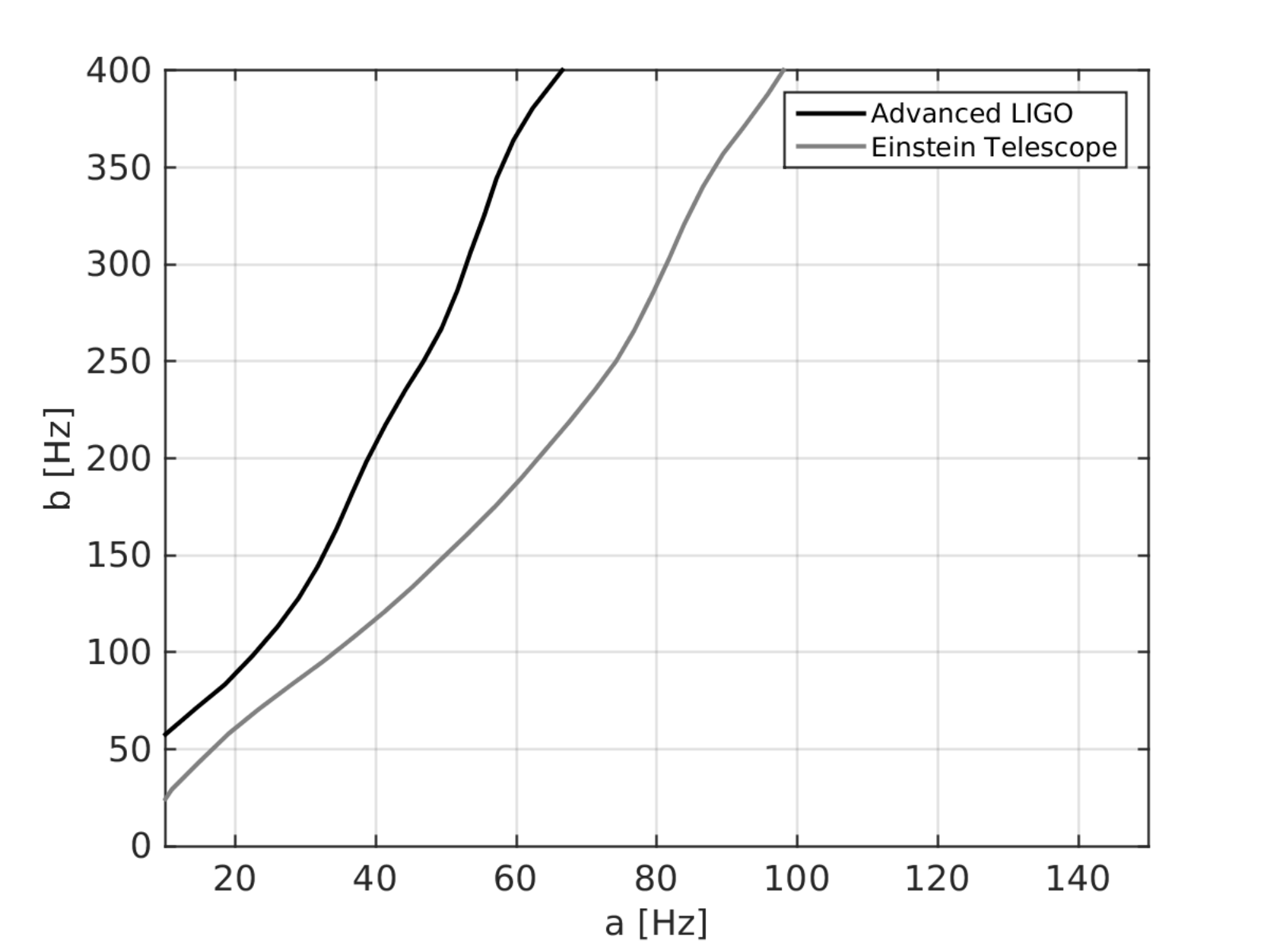} \quad
	\includegraphics[width=3.2in,height=2.5in]{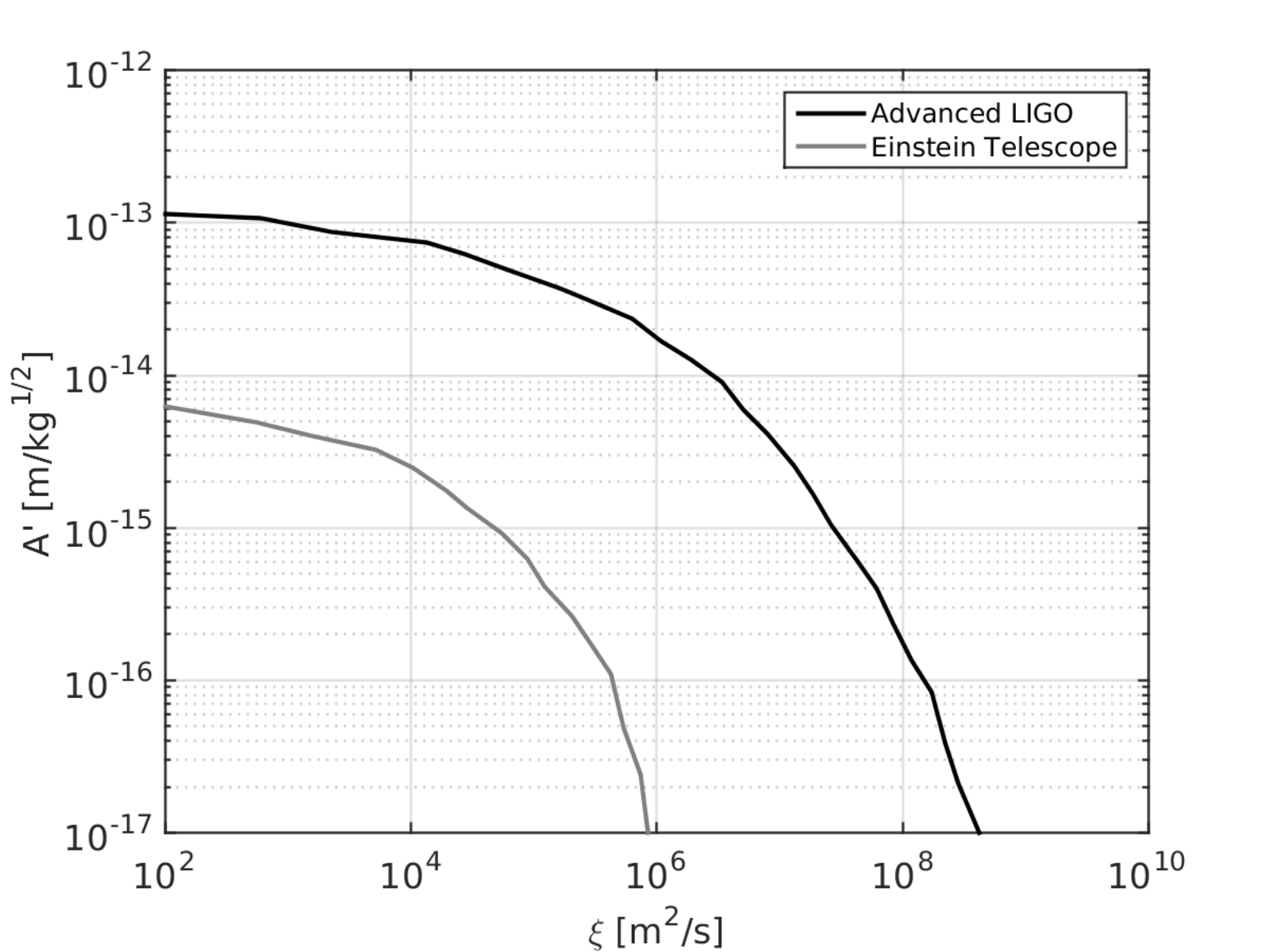} \\
	\end{tabular}
	\caption{Expected 95\% sensitivity contours for Advanced LIGO (black) and Einstein Telescope (gray) in the $b$-$a$ plane (left) and the $A'$-$\xi$ plane (right). The contours in the $\sigma$-$\mu$ plane span most of the area and do not provide significant insight; thus, they are not shown here. More details are provided in the text.}
	\label{combined_model_amplitude}
\end{figure*}

Figure \ref{combined_model_amplitude} shows 95\% sensitivity contours in the $b$-$a$ and $A'$-$\xi$ planes for both Advanced LIGO and the Einstein Telescope, obtained by marginalizing over the other 4 parameters. In the $b$-$a$ plane, the region to the right of the curves is enclosed by the contours; thus, these detectors should be sensitive to parts of the parameter space to the left of their respective contours. This behavior is apparent because $a$ acts as an inverse scaling parameter in our model and $b$ acts similarly to a scaling parameter.

In the $A'$-$\xi$ plane, the region below the curves is enclosed by the contours, so these detectors should be sensitive to areas of the parameter space above the contours. The shape of the contours is immediately apparent since both of the parameters illustrated here are scaling parameters.

The 2D posterior in the $\sigma$-$\mu$ plane spans most of the parameter space and does not yield significant insights into these parameters.

In Table \ref{tab:combined_model_UL}, we compare the expected sensitivities on the $\xi$ and $A'$ parameters for the combined model to those from the individual low- and high-frequency models. The sensitivities for the combined model tend to be better because the combined model includes both low- and high-frequency contributions, increasing the overall GW energy density and improving the detectability of such an SGWB.

\begin{table}[htb!]
  \centering
  \bgroup
  \def\arraystretch{1.2}%
  \begin{tabular}{c  c  c  c  c}
    \hline\hline
    \textbf{Model} & \multicolumn{2}{ c }{$\mathbf{\xi}$ \textbf{[m$^\mathbf{2}$/s]}} & \multicolumn{2}{ c }{\textbf{$\mathbf{A'}$ [m/kg$^\mathbf{1/2}$]}} \\
    & \textbf{aLIGO} & \textbf{ET} & \textbf{aLIGO} & \textbf{ET} \\\hline
    Low frequency & --- & --- & $1.4 \times 10^{-14}$ & $1.4 \times 10^{-15}$ \\\hline
    High frequency & $1.8 \times 10^{11}$ & $3.8 \times 10^{6}$ & --- & --- \\\hline
    Combined & $2.9 \times 10^{7}$ & $2.2 \times 10^{5}$ & $1.1 \times 10^{-14}$ & $1.1 \times 10^{-15}$ \\\hline
    %Low frequency & --- & --- & $1.4 \times 10^{-14}$ & $1.4 \times 10^{-15}$ \\\hline
    %High frequency & $1.8 \times 10^{11}$ & $3.9 \times 10^{6}$ & --- & --- \\\hline
    %Combined & $3.0 \times 10^{7}$ & $2.3 \times 10^{5}$ & $1.1 \times 10^{-14}$ & $1.2 \times 10^{-15}$ \\\hline
    \hline
  \end{tabular}
  \egroup
  \caption[Comparison of sensitivities for $\xi$ and $A'$ for the combined model and the individual low- and high-frequency models.]{Comparison of sensitivities for $\xi$ and $A'$ for the combined model and the individual low- and high-frequency models. The sensitivities for the combined model are better as compared to the individual models due to the overall increase in SGWB energy density for the combined model.}
  \label{tab:combined_model_UL}
\end{table}

\section{Conclusions}\label{sec:conclusions}
In this paper, we have studied the stochastic gravitational wave background generated by stellar core collapse events occurring throughout the universe. Since the energy spectrum of gravitational waves emitted in a single core collapse event is not well understood, we have modeled this spectrum with an empirical functional form, using model parameter ranges determined from fits to a number of numerical simulations of core collapse events. In addition, we have noted that some of the simulations predict a low-frequency peak in the emitted spectrum, possibly due to acoustic waves generated by prompt convection occurring immediately after core bounce. The peak often appears in the vicinity of 100 Hz, and is therefore in or near the most sensitive frequency band for stochastic searches with terrestrial gravitational-wave detectors.

We have performed systematic scans of the model parameter space, including both the empirical broadband spectrum and the low-frequency peak, and compared the resulting stochastic background spectra to the expected sensitivities of the second- and third-generation detectors. While this background is unlikely to be detected by the second-generation, advanced detectors, Einstein Telescope may be able to detect the background in optimistic scenarios. However, in the majority of the parameter space we examined, even the Einstein Telescope would not be sensitive enough to detect this background.

It should be noted that the recent detections of gravitational waves from the binary black hole (BBH) mergers \cite{GW150914,GW151226} have led to new estimates of the stochastic background due to the BBH mergers\cite{GW150914stoch}. The BBH background is estimated to be significantly stronger than the core collapse background for most of the parameter space examined here, and is potentially detectable by advanced detectors. The two backgrounds are predicted to have different spectral shapes (the BBH stochastic background spectrum is proportional to $f^{2/3}$), which offers a potential handle for distinguishing between them. However, distinguishing between the two types of stochastic background will require high signal-to-noise measurement of the stochastic background across the sensitive frequency band 1--100 Hz, as well as a better understanding of the core collapse process and of the gravitational wave spectrum emitted in this process, so that both statistical and systematic uncertainties are smaller than the amplitude of the core collapse background. Further development of numerical simulations of the core collapse is therefore critical for such studies.

\section*{Acknowledgments}
The work K.~A.~O. was supported in part by DOE grant DE-SC0011842 at the University of Minnesota. The work of K.~C., T.~P., and V.~M. was supported in part by NSF grant PHY-1204944 at the University of Minnesota. The work of E.~V. has been carried out at the ILP LABEX (under reference ANR-10-LABX-63) supported by French state funds managed by the ANR within the Investissements d'Avenir programme under reference ANR-11-IDEX-0004-02. LIGO-P1600324.
\bibliography{ccbh}

\end{document}